%% file: main.tex
\documentclass[sigconf]{acmart}
\renewcommand\footnotetextcopyrightpermission[1]{} 
\setcopyright{none}
\renewcommand\footnotetextcopyrightpermission[1]{}

\usepackage[normalem]{ulem}
\usepackage{acronym}
\usepackage{xspace}
\usepackage{comment}
\usepackage{color}
\usepackage{array}
\usepackage{algorithm}
\usepackage[noend]{algpseudocode}
\usepackage{amsmath}
\usepackage{amssymb}
\usepackage[T1]{fontenc}
\usepackage{balance}

\usepackage{bm}
\usepackage{caption}
\usepackage{subfig}
\usepackage{url}
\usepackage{pifont}
\usepackage{wasysym}
\usepackage{fancybox}
\usepackage[most]{tcolorbox}

\input{macro}
\begin{document}
\title{Confidential Inference via Ternary Model Partitioning}
\input{author}
\input{abstract}
\maketitle

\input{introduction}

\input{model}
\input{disclosure}

\input{part}

\input{design}

\input{implementation}

\input{evaluation}

\input{relate}

\input{conclusion}


\bibliographystyle{ACM-Reference-Format}
\bibliography{main}
\end{document}

%% file: macro.tex
\newcounter{TODO}

\newcommand{\nip}[1]{\vspace{1ex}\noindent\textbf{#1}}
\newcommand{\eg}{\emph{e.g.,}\xspace}
\newcommand{\ie}{\emph{i.e.,}\xspace}
\newcommand{\etal}{\emph{et al.}\xspace}

\newcommand{\sysname}{{\textsc{Infenclave}}}
\newcommand{\tmp}{{\textsc{Ternary Model Partitioning}}}
\newcommand{\maf}{{\textsc{Model Assessment Framework}}}

\acrodef{AI}{artificial intelligence}
\acrodef{ML}{machine learning}
\acrodef{MLaaS}{machine learning as a service}
\acrodef{DL}{deep learning}
\acrodef{DNN}{deep neural network}
\acrodef{ConvNet}{convolutional neural network}
\acrodef{EXIF}{Exchangeable Image File Format}
\acrodef{IR}{intermediate representation}
\acrodef{SGX}{Software Guard Extensions}
\acrodef{AES-GCM}{Galois Counter Mode}
\acrodef{TEE}{Trusted Execution Environment}
\acrodef{FLOP}{floating point operation}
\acrodef{KL}{Kullback-Leibler}
\acrodef{MEE}{Memory Encryption Engine}
\acrodef{TEE}{trusted execution environment}
\acrodef{IRGenNet}{IR Generation Network}
\acrodef{IRValNet}{IR Validation Network}
\acrodef{SMC}{secure multi-party computation}
\acrodef{FHE}{fully homomorphic encryption}

\newcommand*\Let[2]{\State #1 $\gets$ #2}
\algrenewcommand\algorithmicrequire{\textbf{Precondition:}}
\algrenewcommand\algorithmicensure{\textbf{Postcondition:}}
\def\HiLi{\leavevmode\rlap{\hbox to \hsize{\color{gray!50}\leaders\hrule height .8\baselineskip depth .5ex\hfill}}}

\acmConference[]{}{}

\usepackage{afterpage}
\usepackage{multicol}
\newsavebox{\shortpagebox}
\makeatletter
\newcommand{\shortpage}[1]
{\par
  \setbox\shortpagebox=\vbox{\strut #1\par}%
  \afterpage{\onecolumn
    \begin{multicols}{2}
    \unvbox\AP@partial
    \end{multicols}}%
  \unvbox\shortpagebox
\par}
\makeatother

%% file: author.tex
\author{Zhongshu Gu}
\affiliation{IBM Research}
\author{Heqing Huang}
\affiliation{ByteDance}
\author{Jialong Zhang}
\affiliation{ByteDance}
\author{Dong Su}
\affiliation{Alibaba}
\author{Hani Jamjoom}
\affiliation{IBM Research}
\author{Ankita Lamba}
\affiliation{Sonatype}
\author{Dimitrios Pendarakis}
\affiliation{IBM Cognitive Systems}
\author{Ian Molloy}
\affiliation{IBM Research}

%% file: abstract.tex
\begin{abstract}
Today's cloud vendors are competing to provide various offerings to simplify and accelerate AI service deployment.
However, cloud users always have concerns about the confidentiality of their runtime data, which are supposed to be processed on third-party's compute infrastructures.
Information disclosure of user-supplied data may jeopardize users' privacy and breach increasingly stringent data protection regulations.
In this paper, we systematically investigate the life cycles of inference inputs in deep learning image classification pipelines and understand how the information could be leaked. 
Based on the discovered insights, we develop a \tmp{} mechanism and bring trusted execution environments to mitigate the identified information leakages.
Our research prototype consists of two co-operative components: (1) \maf{}, a local model evaluation and partitioning tool that assists cloud users in deployment preparation; (2) \sysname{}, an enclave-based model serving system for online confidential inference in the cloud. 
We have conducted comprehensive security and performance evaluation on three representative ImageNet-level deep learning models with different network depths and architectural complexity. 
Our results demonstrate the feasibility of launching confidential inference services in the cloud with maximized confidentiality guarantees and low performance costs.
\end{abstract}

%% file: introduction.tex
\section{Introduction}
\label{sec:introduction}

It has never been easier than today to establish online \ac{AI} services in the public cloud. 
Users can deploy their pre-trained \ac{ML} models to cloud machines and launch services instantly.
However, running services on third-party's infrastructures, but without full control of the computing stacks, continues to pose security and privacy challenges.
Although users expect cloud providers to be trustworthy and
dependable, they still remain cautious about the confidentiality of their supplied data.  
Accidental disclosures of private user data may violate increasingly stricter data protection regulations and lead to reputation damage of service providers.

To address the data protection problem of deploying models in the public cloud,
researchers proposed approaches based on cryptographic primitives\cite{gilad2016cryptonets, mohassel2017secureml,liu2017oblivious} to enable privacy-preserving predictions. 
Although significant performance improvements have been made, these approaches
are still not practical to be deployed in production environments.
Distributed machine learning has also been proposed to
protect data confidentiality\cite{shokri2015privacy,ossia2017hybrid,li2017privynet,mcmahan2016communication}. 
There, part of the model functionality is delegated to the clients
and private data are retained on client machines.  
However, these approaches introduced additional complexities to the client-side program logic.
They also required more computing power on client devices, which are typically resource constrained.

As an alternative, Ohrimenko \etal{} \cite{ohrimenko2016oblivious}
presented data oblivious multi-party machine learning algorithms and
leveraged Intel \ac{SGX} to make them privacy-preserving. 
Recent research efforts, such as \emph{Chiron}\cite{hunt2018chiron} and \emph{Myelin}\cite{hynes2018efficient}, aimed to enable \ac{SGX} enclave integration for \ac{MLaaS}.  
However, \ac{SGX}-based computation is currently performance and
capacity constrained. Specifically, in-enclave workloads cannot exploit external hardware accelerators for computational parallelism.  
\ac{SGX} also has a hard limit (128 MB\footnote{With memory paging support for the Linux \ac{SGX} kernel driver, the virtual memory size of an enclave can be expanded with memory swapping. But swapping on the encrypted memory will significantly affect the performance.})
on the protected physical memory size. 
While, state-of-the-art \ac{DL} models have become increasingly large and complicated despite the limited enclave memory cap.
This, in turn, makes \ac{SGX} inadequate to enclose a large \ac{DL} model entirely within a single enclave.

To address the performance and capacity limitations of secure enclaves, 
it is essential to partition \ac{DL} workloads and delegate computation to run out of enclaves.  
However, \emph{insecure} model partitioning strategies may lead to information disclosure of the original inputs. 
Adversaries may leverage the out-of-enclave \ac{DL} workloads, including both the partial model parameters and the computed outputs, to uncover the properties or even reconstruct the inputs.

In this paper, we systematically investigate the life cycles of 
input data in \ac{DL} image classifiers and explore how adversaries can obtain sensitive information of the inputs. 
Based on the discovered insights, we have derived a \tmp{} mechanism to tightly control deep learning inference pipelines and minimize the information leakage surfaces.
Following the security principles of the \tmp{} mechanism, our research prototype consists of two co-operative components:

(1) On the user side, we have built an offline \maf{} to automate the process of evaluating and partitioning users' proprietary \ac{DL} models from the information-theoretic perspective.
Users can leverage this framework to conduct local model partitioning before deploying their models to the cloud. 

(2) On the cloud side, we have developed \sysname{}, an enclave-based model serving system to host user-supplied models and instantiate online inference services. 
We leverage Intel \ac{SGX} to enable isolated execution with memory encryption protection. 
The sensitive \ac{DL} workload partitions, which are automatically identified and prepared by the \maf{}, would be securely enclosed within the enclave. 
We also enforce authenticated encryption to check the legitimacy of data sources. 
Therefore, we can effectively prevent adversaries from exploiting the \emph{input reconstruction}~\cite{mahendran2015understanding,dosovitskiy2015inverting} and the \emph{model interpretation}~\cite{simonyan2013deep,springenberg2014striving,bach2015pixel,selvaraju2017grad} techniques to divulge sensitive information of the original inputs.
At the same time, we can still delegate non-sensitive workloads to run out of enclave and benefit from hardware-assisted \ac{ML} acceleration. 

Furthermore, our \tmp{} mechanism exploits the layered structure of \acp{DNN} and partitions inference pipelines vertically into multiple components with no functional dependencies. This design aims to leverage the scalability of cloud computing to accommodate higher loads of service requests. To deploy \sysname{} in the cloud, these independent functional components, which hold the sensitive and non-sensitive workloads separately, can be delegated to different physical nodes and horizontally scaled out to serve concurrent requests in parallel. Thus, the capacity constrained enclaves would no longer become the performance bottlenecks of confidential inference services.     

We have conducted comprehensive security and performance evaluation on three representative ImageNet-level \acp{DNN} with different network depths and architectural complexity, \eg{} from the \emph{Darknet Reference Model} (17
layers), \emph{Extraction Model} (28 layers), to the deeper and more complex \emph{DenseNet Model} (306 layers)\cite{huang2017densely}.
The results demonstrate that our system can help users securely and automatically partition their models and enable confidential inference services in the cloud with low performance overhead.

\nip{Roadmap.} 
We present the threat model in Section~\ref{sec:model} to discuss the adversaries we tend to defend against and their capabilities.
Then we analyze the information leakages of \ac{DL} inference pipelines in Section~\ref{sec:leakage}. 
We discuss the security principles of the \tmp{} mechanism in Section~\ref{sec:partition} and describe the system design of the \maf{} and \sysname{} in Section~\ref{sec:design}.   
The evaluation is in Section~\ref{sec:evaluation}, which contains security and performance analysis of our research prototype.
We survey the related works in Section~\ref{sec:relate} and conclude in Section~\ref{sec:conclusion}.

%% file: model.tex
\section{Threat Model}
\label{sec:model}
In our threat model, the goal of the adversaries is to uncover the contents of the user inputs submitted to the cloud-based \ac{DL} inference services.
We consider that the adversaries have the access the cloud machines that serve \ac{DL} models. 
This can be achieved in multiple ways.
For example, adversaries may exploit zero-day vulnerabilities to
penetrate and compromise the system software of the cloud server.  
Insiders, such as cloud administrators, can also retrieve and leak data from the servers on purpose.  
Data can be in the form of files on disks or snapshots of physical memory.  
We assume that adversaries understand the format of the files stored on disks and they are able to locate and extract structured data from memory snapshots.

We assume that cloud users trust \ac{SGX}-enabled processor packages and adversaries cannot break into the perimeters of CPU packages to track the code execution and data flow at the processor level.  
We do not intend to address the side channel attacks against Intel \ac{SGX} in this paper. 
We expect that \ac{SGX} firmware has been properly upgraded to patch recently disclosed micro-architectural vulnerabilities\cite{vanbulck2018foreshadow,chen2018sgxpectre,van2019ridl,schwarz2019zombieload}, and the in-enclave code has been examined to be resilient to side channel attacks.

We assume that the \ac{DL} models to be deployed in the cloud are trained in a secured environment and the model parameters are not leaked to adversaries during model training. 
In addition, we consider that users' devices that submit inference requests are not compromised by adversaries. 
Securing training process\cite{gu2019reaching} and protecting end-point user devices are out of the scope of this paper. 

%% file: disclosure.tex
\section{Information Leakage Analysis}
\label{sec:leakage}
\begin{figure*}[!ht]
\centering
\includegraphics[width=1\textwidth]{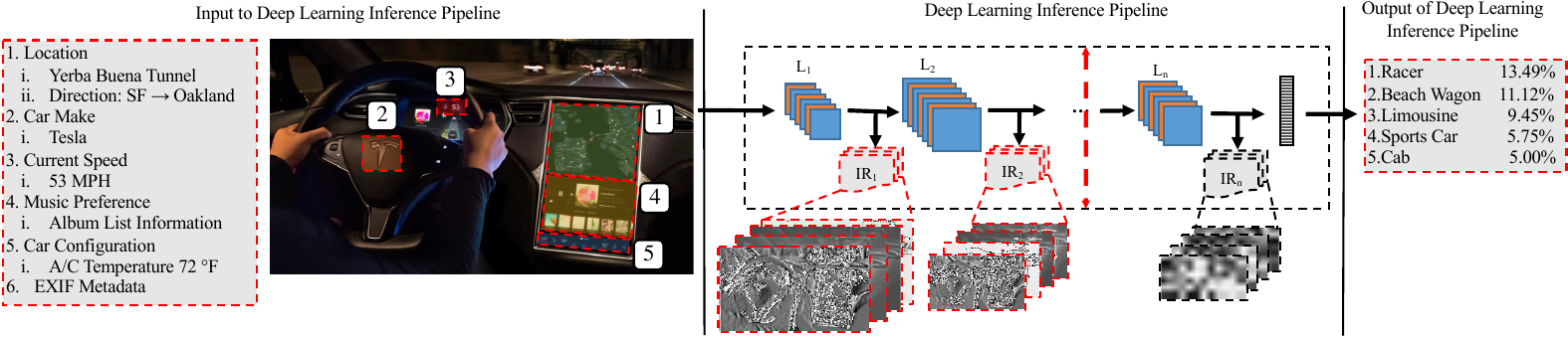}
\caption{The Information Leakages of Input Data in a Deep Learning Inference Pipeline}
\label{fig:privacy}
\end{figure*}

We investigate the life cycles of input data in the \ac{DL} image classification pipelines and intend to identify the potential places where information might leak. 
We give a motivating example in
Figure~\ref{fig:privacy} to empirically demonstrate such information
disclosure.  We feed this picture\footnote{Image source: \url{https://www.tesla.com/software}} into a 1000-class image
classification system. This classifier is powered by a \ac{ConvNet} trained on the ImageNet dataset.  
The output of the system is the top-5 class prediction scores of the picture. 

We find that information leakages of user inputs might happen at multiple stages in a \ac{DL} inference pipeline, \eg{} data entrance, feature extraction, and label mapping. If an adversary is equipped with proper capabilities, he can reconstruct or infer sensitive attributions of the inputs. We describe our discoveries respectively based on the leakage sources in three categories: (i) from \emph{inputs}, (ii) from \emph{\acp{IR}}, and (iii) from \emph{semantic class mapping}.

\subsection{Information Leakages from Inputs}
If the inputs are fed into a \ac{DL} inference system in \emph{unencrypted} formats, it is obvious that adversaries can learn the information directly if they have the access to the inputs.  
For this specific example, adversaries can infer the location of this car based on the map and GPS information \ovalbox{\footnotesize 1}.
The night mode of the map uncovers the approximate time when this picture was taken.  
In addition, the steering wheel emblem \ovalbox{\footnotesize 2}
reveals the car model and the digital speedometer \ovalbox{\footnotesize 3} shows the speed.
Adversaries can also obtain more personal information about the driver from the
album list \ovalbox{\footnotesize 4} of the media player and the car settings \ovalbox{\footnotesize 5} on the touchscreen.  
Furthermore, if the picture's EXIF meta-data are not carefully
eliminated, adversaries may also retrieve the GPS coordinates, the original
date and time when the picture was taken, the device type, and all
camera configurations, which may disclose more private
information about the user. 

\subsection{Information Leakages from \acp{IR}}
First, let us formally define the processes of \ac{DL} inference process and \ac{IR} generation. 
A deep learning inference procedure extracts feature representations layer by layer.  
We can formulate the procedure as a composite transformation function that maps raw inputs to outputs, \ie{} a representation function $F^*$ that maps an input $\mathbf{x}$ to an output $\mathbf{y}$, \ie{} $\mathbf{y} =
F^*(\mathbf{x};\mathbf{\theta})$. $\mathbf{\theta}$ represents parameters that are learned during model training. $F^*$ is composed of $n$ (assuming the network has
$n$ layers) sub-functions $F_i$, where $i \in [1, n]$. $F_i$ maps the
input $\mathbf{x}_i$ to the output $\mathbf{y}_i$ on layer $i$.  These
sub-functions are connected in a chain. Thus, $\mathbf{y} =
F^*(\mathbf{x};\mathbf{\theta}) = \mathit{F}_n
\mathit{F}_{n-1}...\mathit{F}_1(\mathbf{x})$.

Each hidden layer performs its own transformation as a sub-function and generates an \ac{IR} as an output. For a specific layer $p \in [1, n)$, 
$\mathrm{IR} = \Phi(\mathbf{x};
\mathbf{\theta}_{1 \rightarrow p})= \mathit{F}_p
\mathit{F}_{p-1}...\mathit{F}_1(\mathbf{x})$, wherein 
$\mathbf{\theta}_{1 \rightarrow p}$ represents the model parameters of the first $p$ layers and the output $\mathrm{IR}$ is the intermediate representation computed out of this layer. 
A transformation at each hidden layer helps converge the \acp{IR} towards the final outputs.
Based on the research efforts in understanding the internal mechanisms
of \acp{DNN}\cite{zeiler2014visualizing,yosinski2015understanding,shwartz2017opening}, for an image classification \ac{DNN}, the shallow layers respond more to low-level
photographic information (such as edges, corners, and contours, of the
original inputs). In contrast, deep layers represent more abstract and
class-specific information related to the final outputs.
Mapping these insights to the security domain, \acp{IR} computed out of shallow and deep layers may disclose different input information depending on adversaries' capability. 
We summarize three types of information leakages in \acp{IR} as follows:

\paragraph{(1) Input Content Leakages via Projected \acp{IR}.}
\acp{IR} computed out of shallow layers still bestow low-level photographic information of the original inputs. Thus, it is straightforward for humans, acting as adversaries, to understand the \acp{IR} by projecting them to pixel space. 
For example, in Figure~\ref{fig:privacy},
by examining the contents of the \ac{IR} images out of shallow layers, adversaries are able to collect comparable amount of information (though with a certain level of information loss) with regard to the original input. 
We consider that such \acp{IR} at shallow layers explicitly disclose the visual information of the original inputs. 

By progressing towards deeper layers, the projected IR images in pixel space are not \emph{directly} comprehensible. 
The information of the original inputs is transformed into high-level features and is encoded by preceding layers. 
However, the information is still preserved within \acp{IR} and is crucial for final classification decisions. 
If adversaries can obtain both \acp{IR} and \emph{the preceding model parameters}, they can still extract sensitive information of the original inputs by exploiting the \emph{input reconstruction}\cite{mahendran2015understanding,dosovitskiy2015inverting} or the \emph{model interpretation}\cite{simonyan2013deep,springenberg2014striving,bach2015pixel,selvaraju2017grad} techniques. 

\paragraph{(2) Input Content Leakages via Input Reconstruction.}
\label{para:reconstruct}
We consider that adversaries intend to reconstruct the original inputs from the \acp{IR}. 
The input reconstruction problem in \acp{DNN} can be defined as follows: 
Given an $\mathrm{IR} = \Phi(\mathbf{x};
\mathbf{\theta}_{1 \rightarrow p})$, wherein $\mathbf{\theta}_{1 \rightarrow p}$ still represents the model parameters of the first $p$ layers. 
Adversaries tend to compute an approximated inverse function $\phi^{-1}$ to generate $\tilde{\mathbf{x}} = \phi^{-1}(\mathrm{IR})$ that minimizes the distance between $\mathbf{x}$ and $\tilde{\mathbf{x}}$. In practice, in order to derive $\phi^{-1}$, adversaries need to have either \emph{white-box}~\cite{mahendran2015understanding} or \emph{black-box}~\cite{dosovitskiy2015inverting} access to $\mathbf{\theta}_{1 \rightarrow p}$. \emph{White-box} access means that adversaries need to retrieve the model parameters $\mathbf{\theta}_{1 \rightarrow p}$, while \emph{black-box} access means that adversaries need to query $\mathbf{\theta}_{1 \rightarrow p}$ and generate input-IR pairs. The pairs can be further utilized to approximate a surrogate inverse model $\phi^{-1}$. 
Therefore, if adversaries can access both \acp{IR} and $\mathbf{\theta}_{1 \rightarrow p}$, they can recover the information of the original inputs by inverting the \acp{IR}. 

\paragraph{(3) Input Attribution Leakages via Model Interpretation.}
Instead of reconstructing the original inputs, adversaries can also infer attributions of the inputs by exploiting the model interpretation techniques, \eg{} Deconv\cite{zeiler2014visualizing}, Guided Back-propagation\cite{springenberg2014striving}, CAM~\cite{zhou2016learning}, Grad-CAM\cite{selvaraju2017grad}, LRP\cite{bach2015pixel}.
The model interpretation methods can help interpret the numerical values in \acp{IR} and bridge the semantic gap between the input attributions and the activated neurons. 
Similar to the information leakages via input reconstruction, model interpretation also requires \emph{white-box} or \emph{black-box} access to the model parameters. 
Once adversaries can establish the connections from the input attributions to specific numerical values in \acp{IR}, they can generalize to infer sensitive attributions from future inputs.

\subsection{Information Leakages from Semantic Class Mapping}
The last place that might disclose the input information is at the exit of a \ac{DL} inference pipeline. 
The semantic class mapping of a classifier leaks categorical information of the inputs. 
From the top-5 prediction results (\eg{} \emph{racer}, \emph{beach wagon},
\emph{limousine}, \emph{sports car}, or \emph{cab}, with
different probability scores in Figure~\ref{fig:privacy}), we can clearly infer that this picture is highly related to a vehicle, without needing to observe the original input.  
Therefore, \acp{IR} can be deciphered via forward propagation if a model's semantic class labels are left unprotected.

%% file: part.tex
\section{Ternary Model Partitioning}
\label{sec:partition}
Based on the investigation in Section~\ref{sec:leakage}, we devise a \tmp{} mechanism to mitigate the information leakages in \ac{DL} inference pipelines. 
The key idea is to partition each model into three functional components, \ie{} a
\emph{FrontNet}, a \emph{BackNet}, and a \emph{Semantic Class Mapping Unit}.
Thereafter, we enforce execution protection via secure enclaves.

\subsection{Partitioning Mechanism}
Following the mathematical definition of the \ac{DNN} inference, we choose a partitioning layer $p$ where $p \in [1,n)$ in an n-layer \ac{DNN}. The \emph{FrontNet} includes layer $1\rightarrow p$ and the \emph{BackNet} includes the following layers.  
The function
for \emph{FrontNet} can be represented as $\Phi$:
$\mathrm{IR} = \Phi(\mathbf{x};
\mathbf{\theta}_{1 \rightarrow p})= \mathit{F}_p
\mathit{F}_{p-1}...\mathit{F}_1(\mathbf{x})$ and its output
$\mathrm{IR}$ is the intermediate representation computed out of a
\emph{FrontNet}.  The function $\lambda$ for a \emph{BackNet} is
$\mathbf{y} = \lambda(\mathrm{IR}; \mathbf{\theta}_{p+1 \rightarrow n}) = \mathit{F}_n
\mathit{F}_{n-1}...\mathit{F}_{p+1}(\mathrm{IR})$, in which
$\mathrm{IR}$ is the input. The final output $\mathbf{y}$ of a \emph{BackNet} is a probability vector. The top-k entries are extracted from $\mathbf{y}$ and are mapped with a set of user-defined class labels $\mathrm{L}$. This label matching process is conducted in the \emph{Semantic Class Mapping Unit}.    

\subsection{Deployment Strategy}
Cloud users first partition their to-be-deployed models locally. They can designate that the \emph{FrontNet} and \emph{Semantic Class Mapping Unit} should be enclosed within a secure enclave in the cloud, whereas the \emph{BackNet} can run out of the enclave to benefit from hardware acceleration. Thus, they can encrypt both the \emph{FrontNet} submodel and labels with their secret keys and provide them to the cloud. The encrypted \emph{FrontNet} and labels are only allowed to be loaded and decrypted after the enclave initialization and remote attestation. At runtime, cloud users can submit encrypted inputs to the online inference service and the inputs are only allowed to be decrypted within the enclave. The semantic class mapping is also conducted within the enclave and the final classification results are sealed before returning back to users. 

Here we want to emphasize another advantage of the \tmp{} mechanism for cloud deployment. The partitioned three functional components, \ie{} a \emph{FrontNet}, a \emph{BackNet}, and a \emph{Semantic Class Mapping Unit}, have no control dependencies between each other and can be delegated to different physical nodes. 
Therefore, we can significantly benefit from the load scalability of cloud computing to serve higher volume of service requests. Based on the needs of the confidential inference service, we can horizontally scale out by replicating and distributing \emph{FrontNet} and \emph{BackNet} to multiple cloud nodes on demand to handle requests in parallel.

\subsection{Leakage Mitigation}
It is crucial to check whether the \tmp{} mechanism can effectively prevent the information leakages defined in Section~\ref{sec:leakage}. First, inputs will not leak information because they have been encrypted before being submitted to the cloud. Enclave's memory encryption mechanism can prevent the information leakage of the original inputs at runtime. 
Second, the model parameters of the \emph{FrontNet}, which generates the IRs that will be passed to the \emph{BackNet}, are enclosed within secure enclaves. Adversaries cannot retrieve the \emph{FrontNet} parameters because \emph{FrontNet} is always encrypted outside of enclaves. Furthermore, we use \ac{AES-GCM} to authenticate encrypted inputs. Thus, we can prevent adversaries from querying the \emph{FrontNet} as a black-box. This authenticated encryption mechanism can effectively eliminate the attack surface for exploiting the \emph{input reconstruction} and \emph{model interpretation} methods.
Third, the \emph{Semantic Class Mapping Unit} is conducted within the boundary of enclaves. Adversaries cannot decipher the semantic meanings of the probability vector as the labels have already been protected. 

By now, the only \emph{missing piece} is whether our design can prevent the \acp{IR} from leaking information. This also determines how many layers we need to include in the \emph{FrontNet}, which is model-specific. To translate the security requirement in the context of \tmp{}, we need to find a partitioning layer satisfying the following property: the \acp{IR} generated after this layer are no longer similar to the original inputs. In order to address this problem, we develop a \maf{} (Section~\ref{ssec:maf}) for users to automatically find the secure partitioning layers for different \ac{DL} model architectures.

%% file: design.tex
\section{System Design}
\begin{figure}[!t]
\centering
\includegraphics[width=0.5\textwidth]{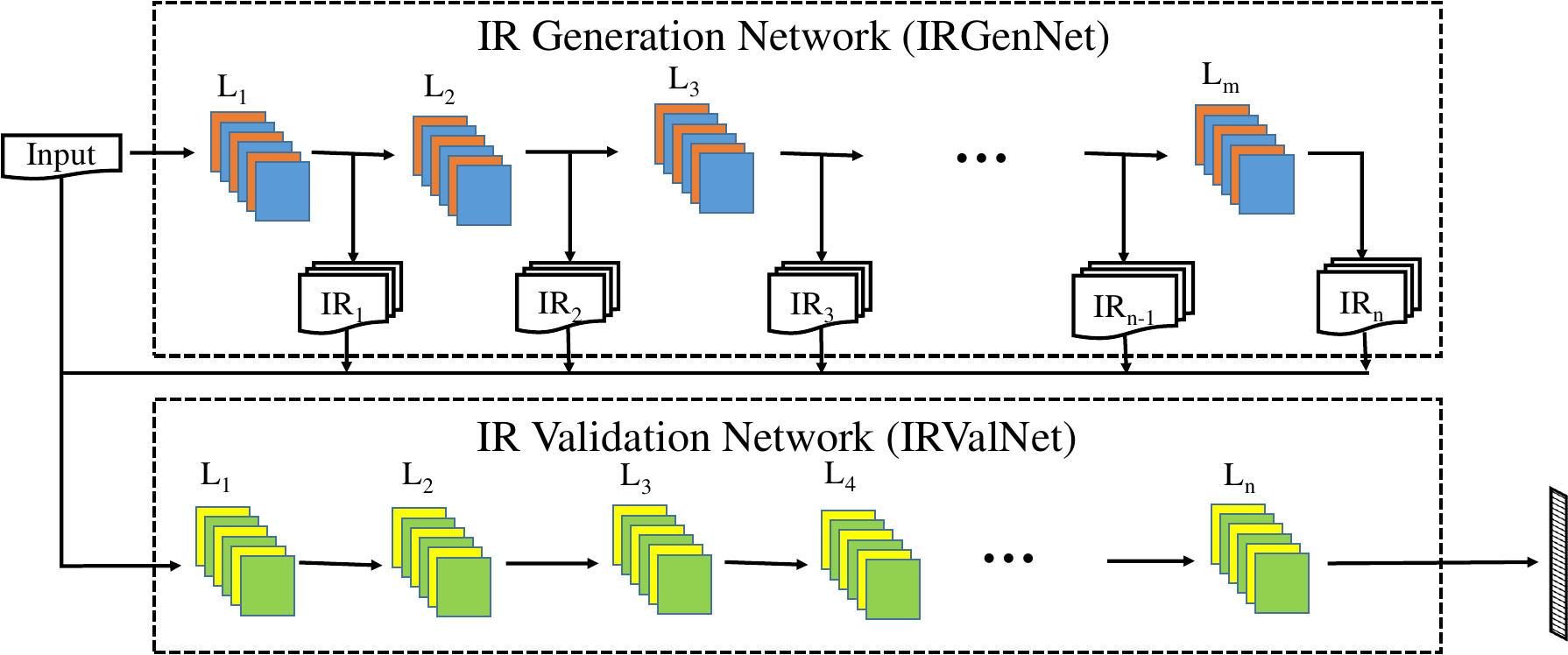}
\caption{Model Assessment Framework}
\label{fig:dnnvalidation}
\end{figure}

\begin{figure*}[!t]
\centering
\includegraphics[width=1\textwidth]{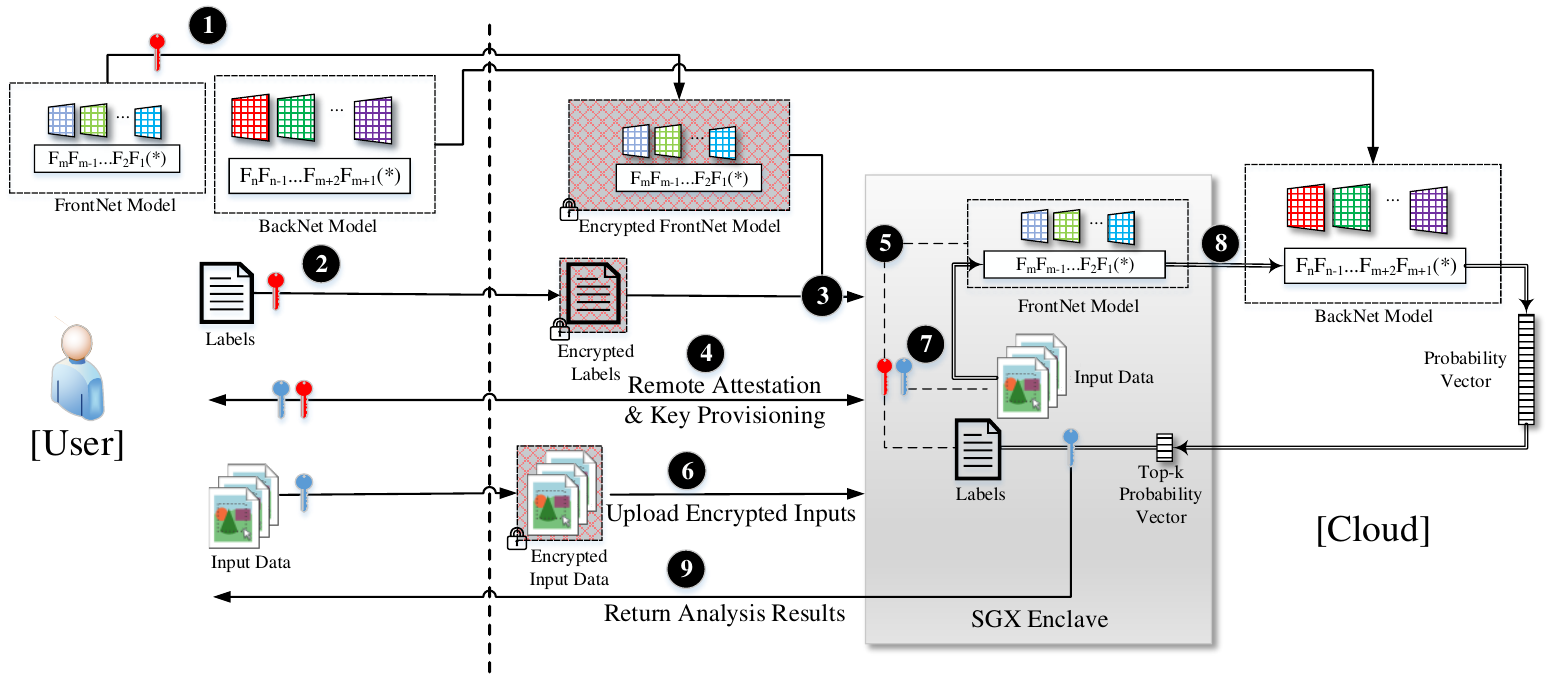}
\caption{The Workflow of the Confidential Inference Service via \sysname{}}
\label{fig:protocol}
\end{figure*}
\label{sec:design}
In this section, we describe the design principles of the two co-operative systems, the offline \maf{} and the online \sysname{}, respectively. 

\subsection{\maf{}}
\label{ssec:maf}

We design the \maf{} based on the key intuition that if \acp{IR} retain similar contents as the original input, they will be classified into similar categories under the same oracle \ac{DNN}.  
By measuring the similarity between the classification probability vectors, we can quantitatively determine whether a specific IR is similar to its original input. 

In Figure~\ref{fig:dnnvalidation}, we present the \emph{Dual-Neural-Network} architecture of our \maf{}. 
We submit an input $\mathbf{x}$ to the \emph{\ac{IRGenNet}} and generate $\mathrm{IR_i}$ for all $i \in \left[1,n\right]$. 
We place the \ac{DL} model that is to be deployed in the public cloud as the \emph{\ac{IRGenNet}} to generate \acp{IR}.
Each $\mathrm{IR_i}$ contains multiple feature maps after passing layer $i$ ($L_i$). 
Then we project feature maps to IR images and submit them to the \emph{\ac{IRValNet}}, which acts as an oracle to inspect IR images. 
The \emph{\ac{IRValNet}} can have different model architecture/weights from the \ac{IRGenNet}.
We use the \emph{DenseNet Model}\cite{huang2017densely}, which is known for its high accuracy in image classification, as the oracle \ac{IRValNet}. 
The output of the \emph{\ac{IRValNet}} is a $N$-dimensional ($N$ is the number of classes) probability vector with class scores.

We define $\mathsf{dist}[\mathbf{x}, \mathrm{IR_i}]$ to quantify the similarity between $\mathbf{x}$ and $\mathrm{IR_i}$. 
We use \ac{KL} divergence to measure the similarity of classification probability distributions for both $\mathbf{x}$ and $\mathrm{IR_i}$.
At each layer $i$, we select the IR image having the minimum \ac{KL} divergence ($D_{KL}$) with the input $\mathbf{x}$: $\forall j \in [1, \mathrm{d(L_i)}]$, where $\mathrm{d(L_i)}$ is the depth of the output tensor of layer $i$.
\begin{equation}
\begin{split}
\mathsf{dist}[\mathbf{x}, \mathrm{IR_i}] &= \emph{min}_j(D_{KL}(F^*(\mathbf{x}, \theta)\ ||\ F^*(\mathrm{IR_{ij}}, \theta))) \\
&= \emph{min}_j(\sum^{N}_{k=1} F^*(\mathbf{x}, \theta)_{[k]} \log \frac{F^*(\mathbf{x}, \theta)_{[k]}}{F^*(\mathrm{IR_{ij}}, \theta)_{[k]}}),
\end{split}
\label{eq:kl}
\end{equation}

where $F^*(\cdot, \theta)$ is the representation function of \emph{\ac{IRValNet}}.  
To determine the secure partitioning layer for each neural network, we also compute $D_{KL}(F^*(\mathbf{x}, \theta) \ ||\ \mu)$ where $\mu \sim \mathtt{U}\,\{1, N\}$, the discrete uniform distribution of the probability vector, and $N$ is the number of classes. This represents that adversaries have no prior knowledge of
$\mathbf{x}$ before obtaining \acp{IR} and consider that $\mathbf{x}$ will be classified to any class with equal chance. We use it as the baseline for comparison. 
Thereafter, we compute
$\forall i \in [1,n], \ \delta_i = \frac{\mathsf{dist}[\mathbf{x}, \mathrm{IR_i}]}{D_{KL}(F^*(\mathbf{x}, \theta)\ ||\ \mu)} $. 
We can choose to partition at layer $i$, \emph{iff} $\forall t \in [i, n], \ \delta_t>1$. 
We inspect all the preceding layers to satisfy this \emph{iff} condition because \ac{KL} divergence scores may fluctuate, especially when encountering some special \ac{DNN} structures, \eg{} \emph{skip connections}. 
We need to ensure that \acp{IR} with observable input information will not be transmitted to \emph{BackNet} layers.
We will elaborate further about this interesting phenomenon in our \emph{DenseNet} case study in Section~\ref{ssec:partanalysis}.

In summary, users can leverage our \maf{} to analyze their deep learning models before deployment and automatically determine the secure \emph{FrontNet}/\emph{BackNet} partitioning. We can ensure that there would be no explicit information leakages in \acp{IR}.

\subsection{Workflow of \sysname{}}
\input{algorithm}
We adopt the \ac{TEE} technology to enable isolated execution for security-sensitive computation of \ac{DL} inference. 
We choose to use Intel \ac{SGX}\cite{mcKeen2013innovative} as the \ac{TEE} in our research prototype, but our approach in principle can also be generalized to other \acp{TEE}\cite{boivie2013secureblue++, kaplan2016amd}. 
With the memory protection mechanisms of \ac{SGX}, all non-enclave accesses from privileged system software will be denied. 
In order to protect the secrecy of user inputs, we enforce enclaved execution in three stages: data entrance, \emph{FrontNet} computation, and semantic class mapping.
Here we describe the design of \sysname{} by explaining the steps of establishing a confidential inference service in Figure~\ref{fig:protocol} and Algorithm~\ref{alg:prediction}.

\nip{\ding{182}} The user first leverages the \maf{} to evaluate and partition a pre-trained proprietary model into a \emph{FrontNet} and a \emph{BackNet}.  
The \emph{FrontNet} is encrypted with the user's symmetric model key.
The user uploads the encrypted \emph{FrontNet} and the plaintext \emph{BackNet} to the \sysname{} system in the cloud. 

\nip{\ding{183}} Similarly, the user encrypts the class label set and uploads the encrypted label set to the cloud. 

\nip{\ding{184}} We initialize an Intel \ac{SGX} enclave (\textsc{init\_sgx} at line 19). After initialization, we securely copy the encrypted \emph{FrontNet} and labels into the enclave (\textsc{sgx\_load\_mdl\_lbl} at line 20).

\nip{\ding{185}}  The user and the SGX enclave need to perform remote attestation\cite{anati2013innovative} before secret provisioning.\footnote{Due to the licensing procedure for registering SGX enclave code and the prerequisites for using the \emph{Intel Attestation Server} (IAS), we currently skip this step and instantiate a TLS session directly between the user and the enclave.}
The detailed description of the standard attestation protocol can be found in an example\cite{sgxremoteattestation} provided by Intel. 
After remote attestation, a secure Transport Layer Security (TLS) communication channel is created and the user provisions symmetric keys (\textsc{sgx\_get\_keys} at line 4) directly into the enclave in the cloud. 

\nip{\ding{186}} Inside the enclave, we verify the integrity of both the model and the labels by checking their \ac{AES-GCM} authentication tags, and decrypt the \emph{FrontNet} model (\textsc{sgx\_verify\_dec} at line 5) and the labels (\textsc{sgx\_verify\_dec} at line 6) with the provisioned symmetric key from the user. Then we can build a \ac{DNN} based on the \emph{FrontNet} (\textsc{sgx\_load\_dnn} at line 7) within the enclave and the \emph{BackNet} (\textsc{load\_dnn} at line 21) out of the enclave.

\nip{\ding{187}} We allow the users to upload their encrypted input data. Similarly, we copy the encrypted input into the enclave and decrypt them after authentication (\textsc{sgx\_verify\_dec} at line 9). 

\nip{\ding{188}} Within the enclave, we pass the decrypted input into the \emph{FrontNet} model (\textsc{sgx\_model\_inf} at line 10) and generate the IR.

\nip{\ding{189}} The generated IR is securely copied out of the enclave through a controlled channel of SGX. We pass the IR into the \emph{BackNet} model and get the probability vector (\textsc{model\_inf} at line 25) for data prediction. 
This vector is an $N$-dimensional vector that represents a probability distribution over $N$ different possible classes.  

\nip{\ding{190}} We extract the top-k entries and pass them into the enclave to find the mapping for their semantic class labels (\textsc{sgx\_class\_mapping}
at line 26). The mapped results are encrypted (\textsc{sgx\_enc} at line 14) with the prior symmetric key provisioned by the user. The encrypted result is returned back to the user.

%% file: algorithm.tex
\begin{algorithm}[!t]
  \caption{Partitioned Deep Learning Inference}
  \footnotesize
  \label{alg:prediction}
\begin{tabular}{|lp{2.5in}|}\hline
Input:
    & \textit{fn\_enc} \Comment{Encrypted \emph{FrontNet} sub-model}\\ 
    & \textit{bn} \Comment{\emph{BackNet} sub-model}\\
    & \textit{lbl\_enc} \Comment{Encrypted model labels}\\
    & \textit{img\_enc} \Comment{Encrypted image input}\\
    & \textit{clt} \Comment{Client key provisioning server address} \\ 
    & \textit{k} \Comment{Number of returned predictions} \\
\hline
\end{tabular}
  \begin{algorithmic}[1]
  \State \textbf{\#\#\#\#\#\#\#\#\#\#\#\#\#\# Within SGX Enclave \#\#\#\#\#\#\#\#\#\#\#\#\#\#}
  
  \Function{sgx\_load\_enc\_mdl\_lbl }{\textit{fn\_enc, lbl\_enc, clt}}
  \Let{\textit{tls}}{\textsc{sgx\_attestation }(\textit{quote, clt})}
  \Let{\textit{model\_key}, \textit{img\_key}}{\textsc{sgx\_get\_keys }(\textit{clt, tls})}
   \Let{fn}{\textsc{sgx\_verify\_dec }(\textit{fn\_enc, model\_key})}
   \Let{lbl}{\textsc{sgx\_verify\_dec }(\textit{lbl\_enc, model\_key})}
  \Let{\textit{frontnet}}{\textsc{sgx\_load\_dnn }(\textit{fn})}
  \EndFunction

\Function{sgx\_inf\_enc\_img }{\textit{img\_enc}}
  \Let{\textit{img}}{\textsc{sgx\_verify\_dec }(\textit{img\_enc, img\_key})}
  \Let{\textit{ir}}{\textsc{sgx\_model\_inf }(\textit{frontnet, img})}
  \State \Return{\textit{ir}}
  \EndFunction
  
  \Function{sgx\_class\_mapping }{\textit{pv\_k}}
  \Let{\textit{result}}{\textsc{sgx\_mapping }(\textit{lbl, pv\_k})}
  \Let{\textit{enc\_result}}{\textsc{sgx\_enc }(\textit{img\_key, result})}
  \State \Return{\textit{enc\_result}}
  
  \EndFunction
  \State
  \State \textbf{\#\#\#\#\#\#\#\#\#\#\#\#\#\# Out of SGX Enclave \#\#\#\#\#\#\#\#\#\#\#\#\#\#}

  \Function{initialize\_enclave }{\textit{fn\_enc, bn, lbl\_enc, clt}}
    \Let{\textit{eid}}{\textsc{init\_sgx }()}
    \State {\textsc{sgx\_load\_enc\_mdl\_lbl } (\textit{eid, fn\_enc, lbl\_enc, clt})}
    \Let{\textit{backnet}}{\textsc{load\_dnn }(\textit{bn})}
    \State \Return{\textit{eid, backnet}}
  \EndFunction
  \Function{inf\_enc\_img }{\textit{eid, img\_enc, k, clt}}
      \Let{\textit{ir}}{\textsc{sgx\_inf\_enc\_img } (\textit{eid, img\_enc})}
    
    \Let{\textit{pv}}{\textsc{model\_inf }(\textit{backnet, ir})}
    \Let{\textit{enc\_result}}{\textsc{sgx\_class\_mapping }(\textit{eid, }\textsc{top }(\textit{pv, k}))}
    \State \Return{\textit{enc\_result}}
  \EndFunction

  \end{algorithmic}
\end{algorithm}

%% file: implementation.tex
\subsection{Implementation}
\label{sec:implementation}
We have developed \sysname{} based on \emph{Darknet}\cite{darknet13}, which is an open source neural network implementation in C and CUDA. 
We have implemented the \maf{} to measure the information leakage of IRs at different layers. It can guide users to determine, for each specific deep learning model, the optimal number of layers to include in a \emph{FrontNet} and run within an enclave. 
In addition, we port the code from the \emph{mbedtls-SGX}\cite{mbedtls-SGX}, which is an mbedtls-based implementation of the TLS protocol suite supporting Intel \ac{SGX}, to enable TLS communication for key provisioning in \sysname{}. 
In total, we add 23,333 SLOC in C and 474 SLOC in Python for the system development.

%% file: evaluation.tex
\section{Evaluation}
\label{sec:evaluation}
In this section, we first conduct a security analysis targeting adversaries with different adversarial purposes. 
Thereafter, we leverage the \maf{} to guide the partitioning of three representative ImageNet-based deep learning models. 
For each partitioned model, we also study the workload allocation based on low-level \acp{FLOP} to see the proportion of computation that can be delegated to run out of enclaves. Thus, we can obtain the hardware-agnostic metrics to quantify the performance benefits from external hardware acceleration. 
Finally, we measure the inference performance of \sysname{} on our testbed machine equipped with \ac{SGX}. 

\subsection{Security Analysis}
Here we consider two hypothetical adversaries, $\mathcal{A}_1$ and $\mathcal{A}_2$. They tend to uncover the contents of the original input $\mathbf{x}$ after obtaining $\mathbf{x}$'s \acp{IR} out of the enclave. 
We consider both adversaries have no prior knowledge of input $\mathbf{x}$, but they have different attack strategies: $\mathcal{A}_1$ intends to reconstruct the original inputs from the exposed \acp{IR}; $\mathcal{A}_2$ tries to infer the attribution information belonging to the inputs through \emph{model interpretation} methods. 

\nip{Input Reconstruction Attacks ($\mathcal{A}_1$).}
 Here we qualitatively review two representative \emph{input reconstruction} techniques for \acp{DNN}, analyze the requirements and preconditions for
these research works, and demonstrate that we
can protect the data confidentiality of user inputs from powerful
adversaries equipped with these techniques.

In Mahendran and Vedaldi\cite{mahendran2015understanding}, the authors proposed a gradient descent based approach to reconstructing original inputs by inverting the \acp{IR}. 
Following the formal description of the input reconstruction problem in Section~\ref{sec:leakage}, the objective of their approach is to minimize the loss function, which is the Euclid distance between $\Phi(\mathbf{x})$ and $\mathrm{IR}$. 
Considering that $\Phi$ should not be uniquely invertible, they restrict the inversion by adding a regularizer to enforce natural image priors. 

The research by Dosovitskiy and Brox\cite{dosovitskiy2015inverting} has a similar goal of inverting the \acp{IR} to reconstruct the original inputs. 
The major difference is that they do not manually define the natural image priors, but learn the priors implicitly and generate reconstructed images with an up-convolutional neural network. 
They involve supervised training to build the up-convolutional neural network, in which the input is the intermediate representation $\Phi(\mathbf{x})$ and the target is the input $\mathbf{x}$. 
Thus, $\mathcal{A}_1$ needs to collect the training pairs $\{\Phi(\mathbf{x}), \mathbf{x}\}$ in this case. 

In our design, the \emph{FrontNet} models are encrypted by users and are only allowed to be decrypted inside \ac{SGX} enclaves. 
Assume $\mathcal{A}_1$ tends to use Mahendran and Vedaldi's approach\cite{mahendran2015understanding} for reconstruction, the representation function $\Phi$, which is equivalent to the \emph{FrontNet} in our case, is not available in plaintext out of the enclave. 
Thus, we can prevent $\mathcal{A}_1$ from conducting optimization to compute both $\phi^{-1}$ and $\tilde{\mathbf{x}}$.
In addition, we can also prevent the adversaries from querying the online \emph{FrontNet} as a black-box service. 
The reason is that we use \ac{AES-GCM} to enable authenticated encryption. 
The enclave code can deny illegitimate requests, whose authentication tags cannot be verified correctly with users' symmetric keys.  
Therefore, $\mathcal{A}_1$ cannot generate training pairs that are required by Dosovitskiy and Brox's approach\cite{dosovitskiy2015inverting} to build the up-convolutional neural network. 
Without the up-convolutional neural network, $\mathcal{A}_1$ cannot reconstruct the original inputs either. 

\nip{Input Attribution Inference Attacks ($\mathcal{A}_2$).}
 To create connections between inputs and \acp{IR}, all the \emph{model interpretation} methods\cite{zeiler2014visualizing,springenberg2014striving,bach2015pixel,zhou2016learning,selvaraju2017grad} require back-propagation through the \ac{DL} model. 
In our design, we always keep the \emph{FrontNet} within a secure enclave.
Therefore, we can effectively eliminate the possibility for $\mathcal{A}_2$ to decipher the semantic meaning of these numerical numbers within the \acp{IR}. 
In addition, we also place the final class mapping within the enclave and only send the encrypted prediction results to end users. 
Thus, we also mitigate the information leakage of predicted classes for the original inputs.

\subsection{Partitioning Analysis}
\label{ssec:partanalysis}

\begin{figure*}[!h]
\centering
\subfloat[Layer 1: Conv]{
\includegraphics[width=0.19\textwidth]{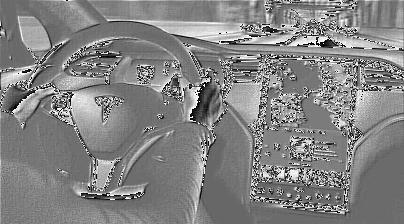}
\label{fig:tesla_darknet_0}}
\subfloat[Layer 2: MaxPool]{
\includegraphics[width=0.19\textwidth]{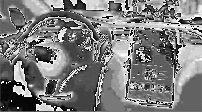}
\label{fig:tesla_darknet_1}}
\subfloat[Layer 3: Conv]{
\includegraphics[width=0.19\textwidth]{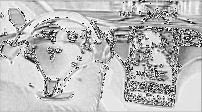}
\label{fig:tesla_darknet_2}}
\subfloat[Layer 4: MaxPool]{
\includegraphics[width=0.19\textwidth]{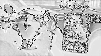}
\label{fig:tesla_darknet_3}}
\subfloat[Layer 5: Conv]{
\includegraphics[width=0.19\textwidth]{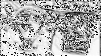}
\label{fig:tesla_darknet_4}}
\caption{The List of IR Images with Minimum KL Divergence at Each Layer (\emph{Darknet Ref. Model})}
\label{fig:tesla_darknet}
\end{figure*}

\begin{figure*}[!h]
\centering
\subfloat[Layer 1: Conv]{
\includegraphics[width=0.19\textwidth]{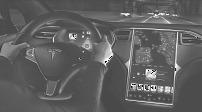}
\label{fig:tesla_extraction_0}}
\subfloat[Layer 2: MaxPool]{
\includegraphics[width=0.19\textwidth]{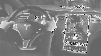}
\label{fig:tesla_extraction_1}}
\subfloat[Layer 3: Conv]{
\includegraphics[width=0.19\textwidth]{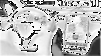}
\label{fig:tesla_extraction_2}}
\subfloat[Layer 4: MaxPool]{
\includegraphics[width=0.19\textwidth]{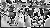}
\label{fig:tesla_extraction_3}}
\subfloat[Layer 5: Conv]{
\includegraphics[width=0.19\textwidth]{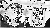}
\label{fig:tesla_extraction_4}}
\caption{The List of IR Images with Minimum KL Divergence at Each Layer (\emph{Extraction Model})}
\label{fig:tesla_extraction}
\end{figure*}

\begin{figure*}[!h]
\centering
\subfloat[Layer 4: Conv]{
\includegraphics[width=0.19\textwidth]{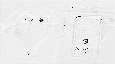}
\label{fig:tesla_densenet_3}}
\subfloat[Layer 5: Route]{
\includegraphics[width=0.19\textwidth]{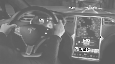}
\label{fig:tesla_densenet_4}}
\subfloat[Layer 21: Conv]{
\includegraphics[width=0.19\textwidth]{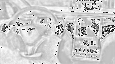}
\label{fig:tesla_densenet_20}}
\subfloat[Layer 59: Conv]{
\includegraphics[width=0.19\textwidth]{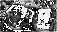}
\label{fig:tesla_densenet_58}}
\subfloat[Layer 205: Conv]{
\includegraphics[width=0.19\textwidth]{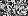}
\label{fig:tesla_densenet_204}}
\caption{The List of IR Images with Minimum KL Divergence at Each Layer (\emph{DenseNet Model})}
\label{fig:tesla_densenet}
\end{figure*}
As an empirical study, we use our \maf{} to determine the secure partitioning layers for three ImageNet-level \acp{DNN}, \ie{} \emph{Darknet Reference Model} (17 layers), \emph{Extraction Model} (28 layers), and \emph{DenseNet Model} (306 layers)\cite{huang2017densely}. 
Based on the partitioning results, we further analyze the workload allocation after partitioning to measure the computation that can be out-sourced to benefit from \ac{ML} accelerators. 
 
\begin{figure}[!t]
\centering
\includegraphics[width=0.5\textwidth]{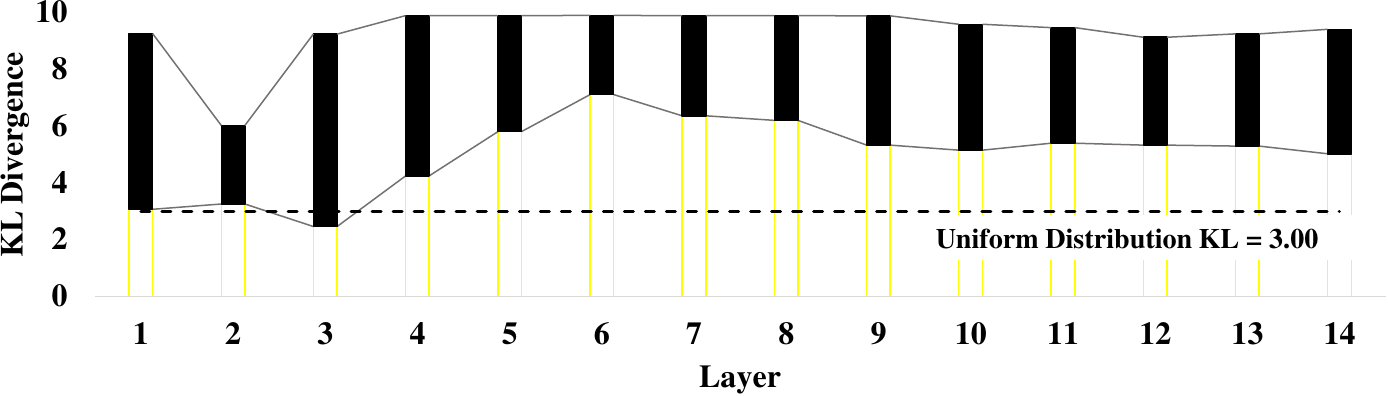}
\caption{KL Divergence for \acp{IR} (\emph{Darknet Ref. Model})}
\label{fig:kl_darknet}
\end{figure}

\begin{figure}[!t]
\centering
\includegraphics[width=0.5\textwidth]{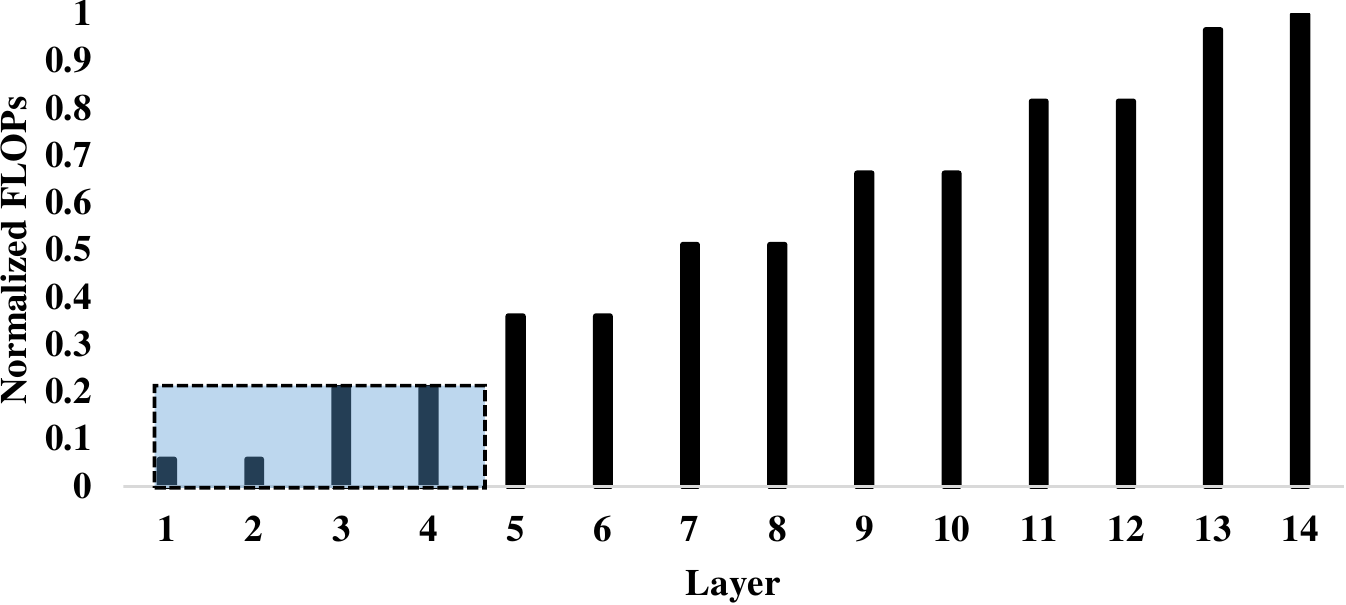}
\caption{Cumulative \acp{FLOP} (\emph{Darknet Ref. Model})}
\label{fig:flop_darknet}
\end{figure}

\nip{Darknet Reference Model.}
This is a relatively small neural network for ImageNet classification. 
Its number of parameters is approximately $1/10$ of \emph{AlexNet}\cite{krizhevsky2012imagenet}, while this model still retains the similar prediction performance (top-1: $61.1\%$ and top-5: $83.0\%$) compared to \emph{AlexNet}. 

To determine the secure partitioning layer for this model, we compare the \ac{KL} divergence ranges of all layers with the discrete uniform distribution. 
In Figure~\ref{fig:kl_darknet}, we present the \ac{KL} ranges (black columns) for the \ac{IR} images of all layers (except the last three layers, \ie{} average pooling, softmax, and cost layers, which do not generate \ac{IR} images).  
For example, at layer $1$ the minimum \ac{KL} divergence is $3.08$ and the maximum is $9.27$. 
The lower the KL divergence score, the more similar it is to the original input. 
We also highlight the line for the \ac{KL} divergence of the discrete uniform distribution with regard to the original input $\mathbf{x}$. This \ac{KL} divergence score is $3.00$.
We can find that after layer $4$, the minimum \ac{KL} divergence scores surpass the line of discrete uniform distribution's \ac{KL}. 
This indicates that exposing \acp{IR} after layer $4$ to adversaries does not reveal more information of the original input compared to any image classified to a uniform distribution. Thus, users can choose to partition the network at layer $4$ and enclose them as a \emph{FrontNet} to run within an enclave.

\begin{figure}[!t]
\centering
\includegraphics[width=0.4\textwidth]{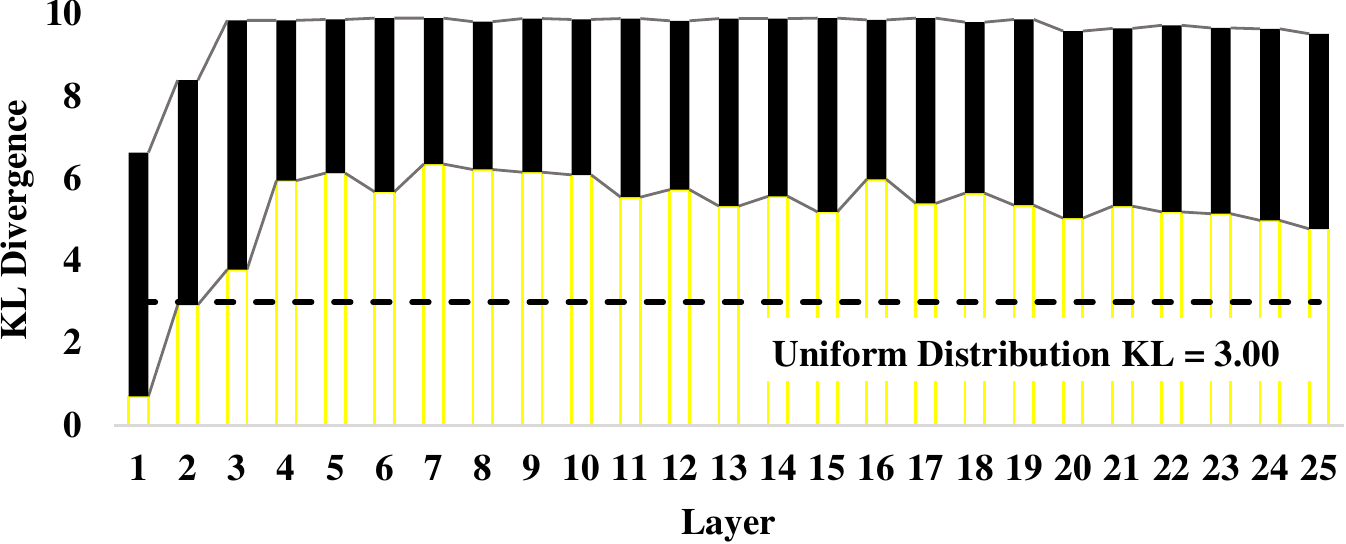}
\caption{KL Divergence for \acp{IR} (\emph{Extraction Model})}
\label{fig:kl_extraction}
\end{figure}

\begin{figure}[!t]
\centering
\includegraphics[width=0.4\textwidth]{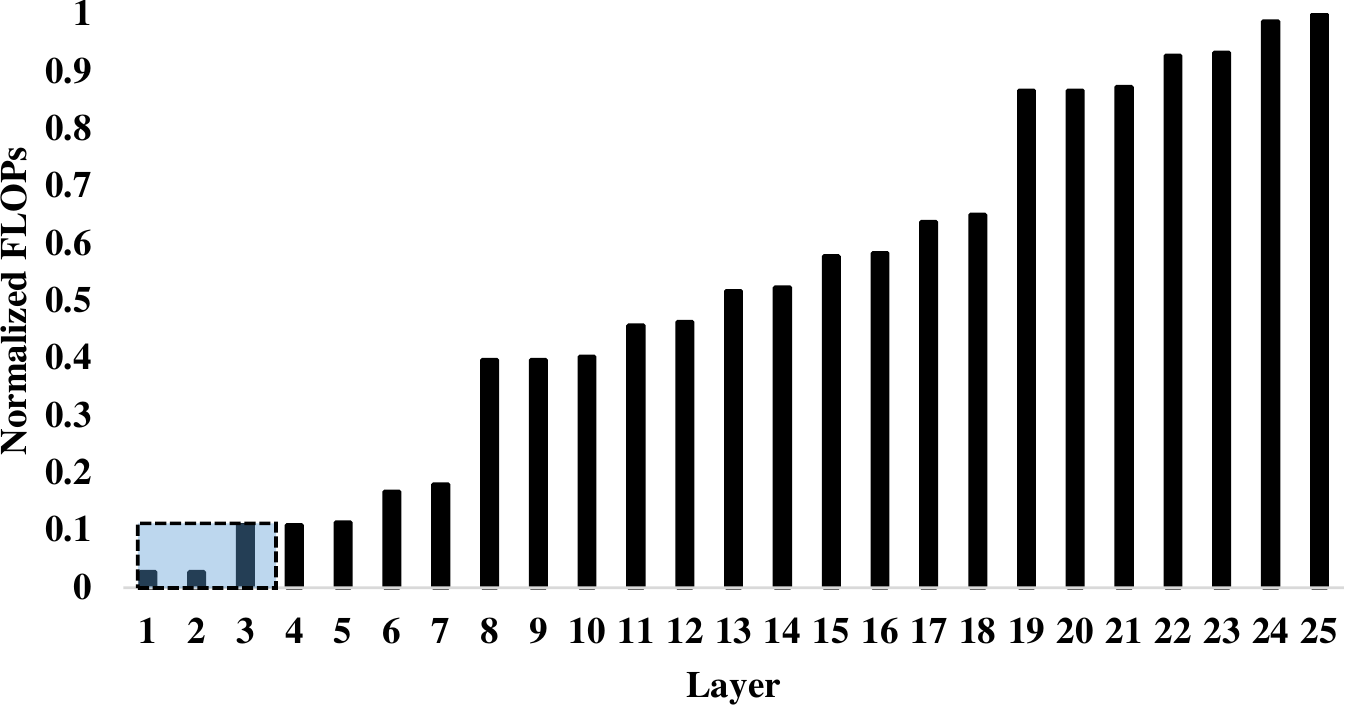}
\caption{Cumulative \acp{FLOP} (\emph{Extraction Model})}
\label{fig:flop_extraction}
\end{figure}

\begin{figure*}[!t]
\centering
\includegraphics[width=1\textwidth]{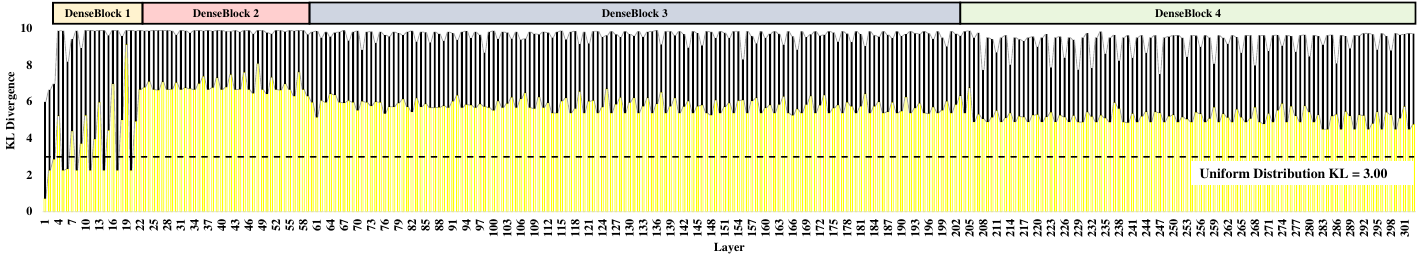}
\caption{KL Divergence for \acp{IR} (\emph{DenseNet Model})}
\label{fig:kl_densenet}
\end{figure*}

\begin{figure*}[!t]
\centering
\includegraphics[width=1\textwidth]{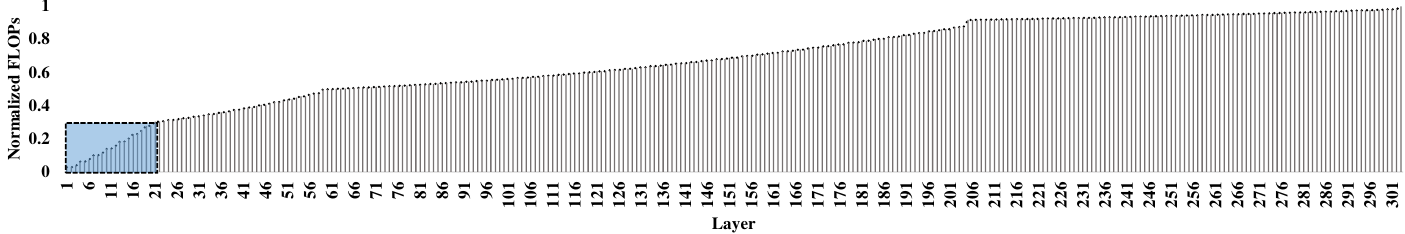}
\caption{Cumulative \acp{FLOP} (\emph{DenseNet Model})}
\label{fig:flop_densenet}
\end{figure*}

To demonstrate that the model assessment result is consistent with human perception, we select the IR images that have the minimum KL divergence scores at the first five layers and display them in Figure~\ref{fig:tesla_darknet}. For example, for layer $1$, the IR image with the minimum KL divergence to the original input is generated by its
7th filter. We can find that, the \ac{IR} images for this specific model retain less and less photographic information of the original inputs when progressing towards deeper layers. From the adversary's perspective, it becomes more difficult to reconstruct the original inputs if he can only obtain \acp{IR} generated by deeper layers running out of enclave.

In addition, we also calculate the workloads of \acp{FLOP} for each layer and display the normalized cumulative \acp{FLOP} in Figure~\ref{fig:flop_darknet}. Based on the partitioning decision above, we can enclose the first four layers into an enclave (shown as the blue box), which comprises only $20.88\%$ of the whole workload. The remaining $79.12\%$ workload can still benefit from out-of-enclave hardware acceleration.      

\nip{Extraction Model.}
Compared to the \emph{Darknet Reference Model}, the \emph{Extraction Model} is deeper and can achieve higher prediction accuracy (top-1: $72.5\%$ and top-5: $90.0\%$). 

We present the \ac{KL} ranges for all layers in Figure~\ref{fig:kl_extraction}.
We can observe a similar phenomenon that after layer $3$, the \ac{KL} divergence score ranges exceed the KL divergence of uniform distribution. Thus the safe partitioning point for this neural network can be at layer $3$.
We also display the \ac{IR} images with the minimum KL
divergence scores for the first five layers in Figure~\ref{fig:tesla_extraction}. 
We can find
that adversaries may still observe residual private information of
the original input from the IR images at layer $1$ and $2$. This is
consistent with the quantitative analysis, which shows a subset of
IR images before layer $3$ have lower KL divergence scores than the
uniform distribution

As demonstrated in Figure~\ref{fig:flop_extraction}, these three-layer enclaved execution only comprises $10.9\%$ of the whole workload.  
The remaining $89.1\%$ workload can be out-sourced to run out of enclaves.  

\nip{DenseNet Model.}
In classical \ac{ConvNet} architectures, each layer obtains the input only from its precedent layer. However, with the increase of network depth, it may lead to the \emph{vanishing gradient problem}\cite{glorot2010understanding,bengio1994learning}. To address this issue, researchers introduced short paths cross layers to make it practical to train very deep neural networks. \emph{DenseNet}\cite{huang2017densely} introduced the neural network topology with \emph{DenseBlocks}. Within each \emph{DenseBlock}, each layer obtains inputs from \emph{all} preceding layers and also transfers its own IRs to \emph{all} subsequent layers. Between two adjacent \emph{DenseBlocks}, it contains \emph{transitional layers} to adjust the \ac{IR}'s size. We find that the information disclosure properties of such special model structures, i.e., \emph{DenseBlocks} and densely connected layers, can be consistently quantified via \ac{KL} divergence analysis. 

We show the \ac{KL} divergence scores of \acp{IR} in Figure~\ref{fig:kl_densenet}. The \emph{DenseNet Model} has four \emph{DenseBlocks}. In each \emph{DenseBlock}, the minimum \ac{KL} divergence scores plummet regularly every two layers.  The reason of this phenomenon is that there exist \emph{route layers} (after every two consecutive convolutional layers) that receive inputs from all preceding layers in the same \emph{DenseBlock}.
For example, the minimum \ac{KL} divergence of layer $4$ (convolutional layer) is $5.24$, while at layer $5$ (\emph{route layer}) it drops to $2.27$. 
Lower \ac{KL} divergence scores indicate higher similarity of two probability distributions.
We can obviously find that layer $5$ (Figure~\ref{fig:tesla_densenet_4}) preserves more information
of the original input than layer $4$ (Figure~\ref{fig:tesla_densenet_3}).
This result implies that we cannot simply partition in the middle of \emph{DenseBlock} 1. 
The IRs generated by deep layers within \emph{DenseBlock} 1 can still reveal original input's information. 
However, there is no densely connected path that \emph{crosses} different \emph{DenseBlocks}. Although there still exist fluctuations of \ac{KL} divergence scores in \emph{DenseBlock} 2, the scores are significantly larger than layers in \emph{DenseBlock} 1. 
In Figure~\ref{fig:tesla_densenet}, we also display the
IR images with minimum KL divergence at all transitional layers
(layer $21$, $59$, and $205$) between different \emph{DenseBlocks}. Based on
the discrete uniform distribution \ac{KL} divergence (3.00), the optimal
partition point is at layer $21$ (the last layer of \emph{DenseBlock} 1).

Similarly, based on Figure~\ref{fig:flop_densenet}, such 21-layer (\ie{} partition at the end of \emph{DenseBlock} 1) \ac{DNN} computation accounts for $30.3\%$ of the whole \ac{FLOP} workload. We can assign the remaining 285 layers to build the \emph{BackNet} and run it out of the enclave.  

\subsection{Performance Analysis}
\label{ssec:evaluation}
In the performance analysis, We evaluated the inference performance of \sysname{} by measuring the average time for passing testing samples through the \emph{Extraction Model}. 
The testing samples were randomly drafted from the \emph{test dataset} of the Tiny ImageNet~\cite{tinyimagenet}.
Our testbed is equipped with an Intel i7-6700 3.40GHz CPU with 8 cores, 16GB of RAM, and running Ubuntu Linux 16.04 with kernel version 4.4.0.

In the baseline case, we loaded the whole \ac{DNN} without using \ac{SGX} and obtained the average time for predicting these unencrypted images. 
To compare with the baseline case, we partitioned the \emph{Extraction Model}, loaded different number of layers (from 1 to 10) as the \emph{FrontNet} inside an \ac{SGX} enclave and the following layers out of the enclave, and obtained the same performance metrics. 
During the performance testing, we launched \sysname{} as an inference service. 
Both the \emph{FrontNet} and the labels were encrypted and they were decrypted within the enclave during the service initialization.  
After the service established, encrypted images were submitted as inputs to \sysname{}. 
The encrypted images were decrypted at runtime inside the enclave. 

\begin{figure}[!t]
\centering
\includegraphics[width=0.5\textwidth]{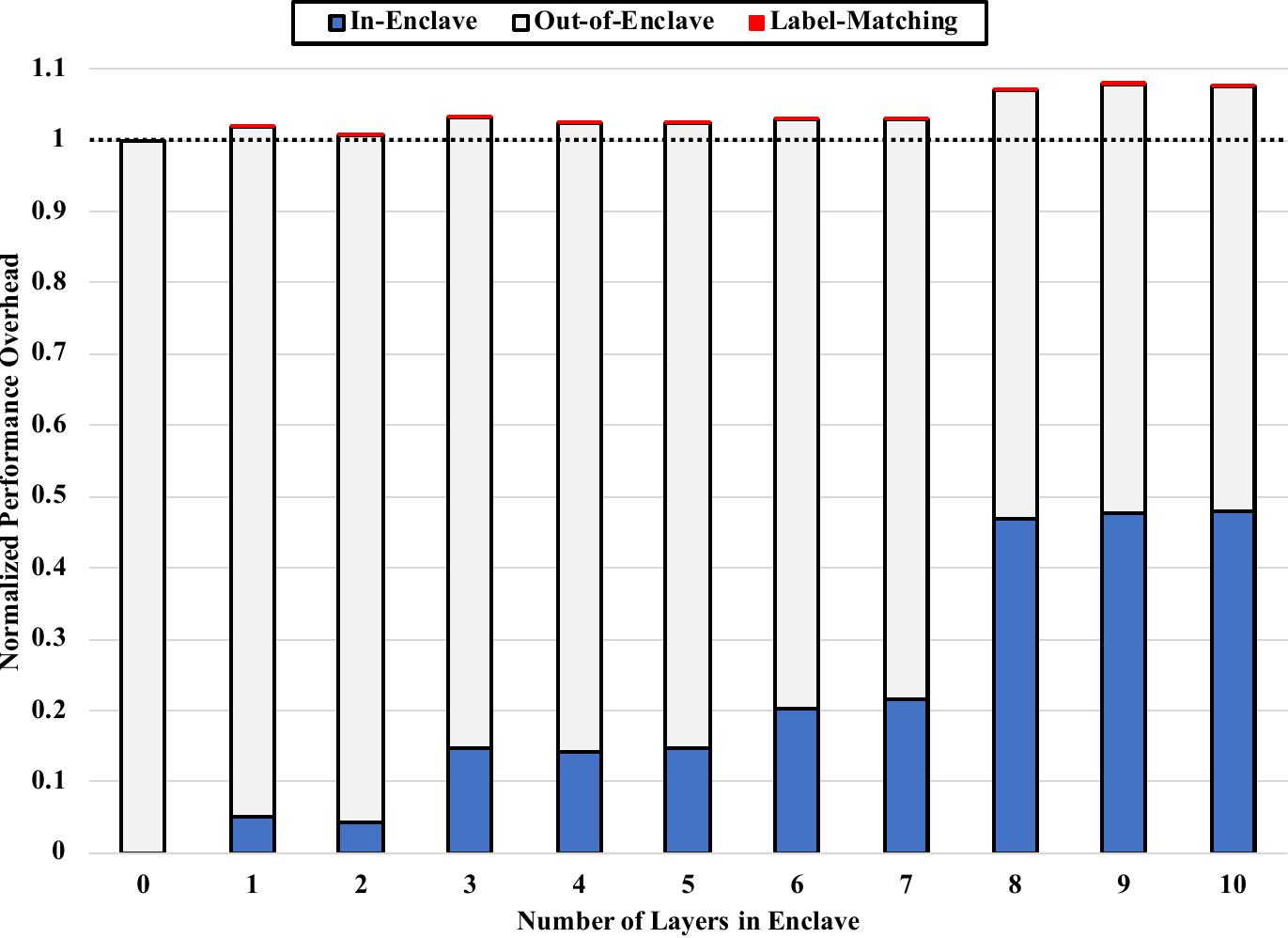}
\caption{Performance Overhead of Running \emph{FrontNet} in an SGX Enclave}
\label{fig:perf_expect}
\end{figure}

\nip{Inference Performance Overhead.} 
We present the normalized performance results in Figure~\ref{fig:perf_expect}. 
We distinguish the performance overhead contributed by in-enclave computation, out-of-enclave computation, and label matching with bars of different colors.
The x-axis shows the number of layers enclosed within the enclave. 
The first column (labeled as ``0'') is the baseline case for unencrypted inputs and all layers run out of the enclave.

First, we can observe that by including more layers into the \emph{FrontNet}, there would be more computation within the enclave and correspondingly less out of enclave.
Second, the performance overhead introduced by the label matching within the enclave is trivial compared to the neural network computation. 
Third, the overall performance overhead increase with more layers within the enclave. 
We can find that when we enclosed 10 layers into the enclave, the performance overhead increased up to 7.5\%. 
This overhead mainly comes from the memory encryption protection of the \ac{SGX} enclave and we consider it is acceptable in the cloud scenario.

\begin{figure}[!t]
\centering
\includegraphics[width=0.5\textwidth]{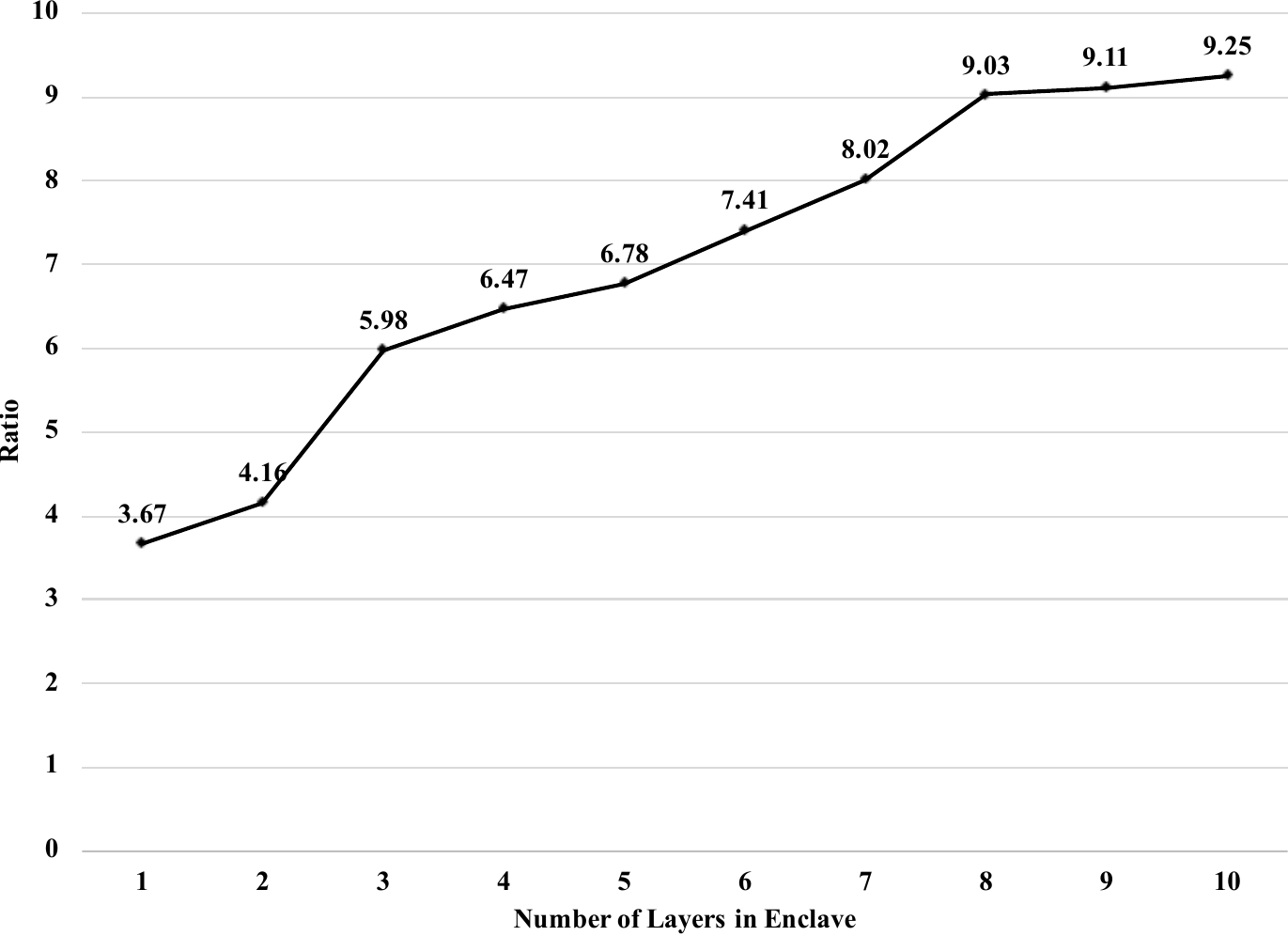}
\caption{Ratio of Processing Time of the First Request to the Second Request}
\label{fig:perf_ratio}
\end{figure}

\nip{A Slow-Start Phenomenon of the Enclaved Service.} 
Here we intend to mention one interesting phenomenon we observed during the performance experiment. 
After launching \sysname{} as an inference service with \ac{SGX} enabled, the time to process the \emph{first} input is much longer than the time to process all the following individual requests. 
We present the ratio of the processing time of the first input to the second input in Figure~\ref{fig:perf_ratio}. 
This ratio increases if we enclose more layers inside the enclave. 
For example, if we run 10 layers inside the enclave, the time to process the first request is 9.25x longer than processing the second request. 
However, this phenomenon did not occur in the baseline case, in which all layers run out of the enclave. 
This is a one-time performance overhead only for the first request and do not affect the performance for the following inputs. 
We speculate that this first-request slow-down phenomenon is highly related to the specific memory loading and CPU cache behaviors of \ac{SGX}. 
To process the first input, the data of the model would be brought from the encrypted memory to the CPU cache, with additional cost of passing through the \ac{SGX} \ac{MEE}.
This explains why processing the first request takes much longer and the time increases if we loaded more layers into the enclave.  
While, the following requests can greatly benefit from the warmed-up cache and significantly decrease the processing time.     

\nip{Performance Remarks.} Compared to cryptographic schemes based approaches\cite{gilad2016cryptonets, mohassel2017secureml, liu2017oblivious} and running whole \ac{DNN} within a single enclave\cite{ohrimenko2016oblivious}, the performance overhead of \sysname{} makes online deep learning inference feasible and adaptable for production-level large-scale \acp{DNN}. 
The out-of-enclave \emph{BackNet} computation can still benefit from hardware acceleration and we grant users the freedom to adjust network partitioning strategy to satisfy their specific security and performance requirements.

%% file: relate.tex
\section{Related Work}
\label{sec:relate}
In this section we list the research efforts that are closely related to our work and highlight our unique contributions compared to these works. 

\nip{Cryptographic Schemes Based Machine Learning.} 
Most of the existing privacy-preserving machine learning solutions are based on cryptographic schemes, such as \ac{SMC}, \ac{FHE}\cite{gentry2009fully}, etc.  
Solutions based on \ac{SMC} protect intermediate results of the computation when multiple parties perform collaborative machine learning on their private inputs.  
\ac{SMC} has been used for several fundamental machine learning tasks\cite{lindell2000privacy, du2004privacy, vaidya2002privacy, vaidya2008privacy, jagannathan2005privacy,mohassel2017secureml}.

Besides these protocol-based solutions, recently researchers also propose to leverage cryptographic primitives to perform deep learning inference.  
Gilad-Bachrach \etal{}\cite{gilad2016cryptonets} proposed \emph{CryptoNets}, a neural network model that makes predictions on data encrypted by \ac{FHE} schemes.  
This approach protects the privacy of each individual input in terms of confidentiality. 
\emph{MiniONN}\cite{liu2017oblivious} is an approach that transforms existing neural networks to an oblivious neural network that supports privacy-preserving predictions. 

Considering the significant performance overhead of using cryptographic schemes, we propose to leverage Intel \ac{SGX} technology to securely execute deep neural network computation on the cloud side. Hence we can protect the confidentiality of user inputs for predictions and can defend against input reconstruction and input attribution inference attacks.     

\nip{Distributed Deep Learning.}
Shokri and Shmatikov\cite{shokri2015privacy} designed a distributed privacy-preserving deep learning protocol by sharing selective parameters and gradients for training deep neural network in a differentially private way.  
Ossia \etal{}\cite{ossia2017hybrid} proposed a distributed machine learning architecture to protect the user's input privacy. Their framework consists of a feature extractor on the mobile client side and a classifier on the server side. 
The server side performs inference task on the dimension-reduced extracted features from the mobile client. 
\emph{PrivyNet}\cite{li2017privynet} is a splitting model deep learning training approach. They reused layers of pre-trained models for feature extraction on local machines and train the cloud neural network for the learning tasks based on the feature representations generated by the local neural network.

Different from their works, our approach leverages \acp{TEE} in the cloud directly to guarantee the confidentiality of user inputs, class labels, and the user-provisioned models. Thus, we significantly simplify the client's logic and relieve client devices, which are supposed to have limited computing capacity and power usage restriction, from heavyweight neural network computation. 
In addition, our approach does not involve transferring intermediate representations through the network, thus eliminating the additional performance overhead for dimension reduction or data compression.

\nip{SGX Applications.}
Intel developed SGX technology to tackle this problem by leveraging the trusted hardware on remote machines. 
A set of new instructions and memory access control have been added since the release of the Intel 6th generation \emph{Skylake} architecture. The general introduction of SGX can be found in \cite{mcKeen2013innovative}. 
We can also find the technical details about the SGX attestation and sealing mechanisms in \cite{anati2013innovative}, dynamic memory allocation of SGX2 in \cite{mckeen2016intel}, and \ac{MEE} in \cite{gueron2016memory}. 
In addition, Costan and Devadas\cite{costan2016intel} gave a detailed explanation and analysis of Intel SGX technology from the perspective of security researchers outside of Intel. 
We have discussed and compared with recent research efforts\cite{ohrimenko2016oblivious, hunt2018chiron, hynes2018efficient} of employing SGX for privacy-preserving machine learning tasks.
To address the performance and capacity limitations of secure enclaves, 
Tram\`er and Boneh\cite{tramer2018slalom} explored a similar workload partitioning methodology. 
They proposed to out-source linear layers' computation of \acp{DNN} to out-of-enclave GPUs. 
Different from their approach, we employ a vertical layer-wise partitioning strategy to exploit the intrinsic structural properties of deep learning models. 
These two partitioning strategies are not in conflict and can be deployed together to further reduce the performance overhead of enclaved inference computation with confidentiality protection.
MLCapsule~\cite{hanzlik2018mlcapsule} is another interesting offline model deployment approach that executes model locally on the client's machine and protects the models' secrecy with \ac{SGX} enclaves. Thus, they explore how to securely deploy server's machine learning workload to the client, while we investigate in a reverse direction on how to out-source client's computation to the server.

%% file: conclusion.tex
\section{Conclusion}
\label{sec:conclusion}
We systematically study the potential information leakages for serving deep learning image classifiers in the public cloud. 
To minimize the attack surfaces of such input data exposure,
we devise a \tmp{} mechanism and leverage trusted execution environments to secure the deep learning inference computation.
On the user side, we design a \maf{} to automate the process of evaluating layer-wise information leakages in deep neural networks and help users determine the optimal model partitioning.
On the cloud side, we build \sysname{}, an enclave-based model serving system, to protect the confidentiality of both user inputs along with user-specified neural network layers and semantic class mapping. 
Our evaluation demonstrates that our system can effectively balance between security and performance to enable confidential deep learning inference in the public cloud. 

%% file: main.bbl

\begin{thebibliography}{48}


\ifx \showCODEN    \undefined \def \showCODEN     #1{\unskip}     \fi
\ifx \showDOI      \undefined \def \showDOI       #1{#1}\fi
\ifx \showISBNx    \undefined \def \showISBNx     #1{\unskip}     \fi
\ifx \showISBNxiii \undefined \def \showISBNxiii  #1{\unskip}     \fi
\ifx \showISSN     \undefined \def \showISSN      #1{\unskip}     \fi
\ifx \showLCCN     \undefined \def \showLCCN      #1{\unskip}     \fi
\ifx \shownote     \undefined \def \shownote      #1{#1}          \fi
\ifx \showarticletitle \undefined \def \showarticletitle #1{#1}   \fi
\ifx \showURL      \undefined \def \showURL       {\relax}        \fi
\providecommand\bibfield[2]{#2}
\providecommand\bibinfo[2]{#2}
\providecommand\natexlab[1]{#1}
\providecommand\showeprint[2][]{arXiv:#2}

\bibitem[\protect\citeauthoryear{Anati, Gueron, Johnson, and Scarlata}{Anati
  et~al\mbox{.}}{2013}]%
        {anati2013innovative}
\bibfield{author}{\bibinfo{person}{Ittai Anati}, \bibinfo{person}{Shay Gueron},
  \bibinfo{person}{Simon Johnson}, {and} \bibinfo{person}{Vincent Scarlata}.}
  \bibinfo{year}{2013}\natexlab{}.
\newblock \showarticletitle{Innovative technology for CPU based attestation and
  sealing}. In \bibinfo{booktitle}{\emph{Proceedings of the 2nd international
  workshop on hardware and architectural support for security and privacy}}.
\newblock


\bibitem[\protect\citeauthoryear{Bach, Binder, Montavon, Klauschen, M{\"u}ller,
  and Samek}{Bach et~al\mbox{.}}{2015}]%
        {bach2015pixel}
\bibfield{author}{\bibinfo{person}{Sebastian Bach}, \bibinfo{person}{Alexander
  Binder}, \bibinfo{person}{Gr{\'e}goire Montavon}, \bibinfo{person}{Frederick
  Klauschen}, \bibinfo{person}{Klaus-Robert M{\"u}ller}, {and}
  \bibinfo{person}{Wojciech Samek}.} \bibinfo{year}{2015}\natexlab{}.
\newblock \showarticletitle{On pixel-wise explanations for non-linear
  classifier decisions by layer-wise relevance propagation}.
\newblock \bibinfo{journal}{\emph{PloS one}} \bibinfo{volume}{10},
  \bibinfo{number}{7} (\bibinfo{year}{2015}), \bibinfo{pages}{e0130140}.
\newblock


\bibitem[\protect\citeauthoryear{Bengio, Simard, and Frasconi}{Bengio
  et~al\mbox{.}}{1994}]%
        {bengio1994learning}
\bibfield{author}{\bibinfo{person}{Yoshua Bengio}, \bibinfo{person}{Patrice
  Simard}, {and} \bibinfo{person}{Paolo Frasconi}.}
  \bibinfo{year}{1994}\natexlab{}.
\newblock \showarticletitle{Learning long-term dependencies with gradient
  descent is difficult}.
\newblock \bibinfo{journal}{\emph{IEEE transactions on neural networks}}
  \bibinfo{volume}{5}, \bibinfo{number}{2} (\bibinfo{year}{1994}),
  \bibinfo{pages}{157--166}.
\newblock


\bibitem[\protect\citeauthoryear{Boivie and Williams}{Boivie and
  Williams}{2013}]%
        {boivie2013secureblue++}
\bibfield{author}{\bibinfo{person}{Rick Boivie} {and} \bibinfo{person}{Peter
  Williams}.} \bibinfo{year}{2013}\natexlab{}.
\newblock \bibinfo{booktitle}{\emph{{SecureBlue++: CPU Support for Secure
  Executables}}}.
\newblock \bibinfo{type}{{T}echnical {R}eport}. \bibinfo{institution}{Research
  report, IBM}.
\newblock


\bibitem[\protect\citeauthoryear{Chen, Chen, Xiao, Zhang, Lin, and Lai}{Chen
  et~al\mbox{.}}{2018}]%
        {chen2018sgxpectre}
\bibfield{author}{\bibinfo{person}{Guoxing Chen}, \bibinfo{person}{Sanchuan
  Chen}, \bibinfo{person}{Yuan Xiao}, \bibinfo{person}{Yinqian Zhang},
  \bibinfo{person}{Zhiqiang Lin}, {and} \bibinfo{person}{Ten~H Lai}.}
  \bibinfo{year}{2018}\natexlab{}.
\newblock \showarticletitle{SGXPECTRE Attacks: Leaking Enclave Secrets via
  Speculative Execution}.
\newblock \bibinfo{journal}{\emph{arXiv preprint arXiv:1802.09085}}
  (\bibinfo{year}{2018}).
\newblock


\bibitem[\protect\citeauthoryear{Costan and Devadas}{Costan and
  Devadas}{2016}]%
        {costan2016intel}
\bibfield{author}{\bibinfo{person}{Victor Costan} {and}
  \bibinfo{person}{Srinivas Devadas}.} \bibinfo{year}{2016}\natexlab{}.
\newblock \showarticletitle{Intel SGX Explained.}
\newblock \bibinfo{journal}{\emph{IACR Cryptology ePrint Archive}}
  (\bibinfo{year}{2016}), \bibinfo{pages}{86}.
\newblock


\bibitem[\protect\citeauthoryear{Dosovitskiy and Brox}{Dosovitskiy and
  Brox}{2016}]%
        {dosovitskiy2015inverting}
\bibfield{author}{\bibinfo{person}{Alexey Dosovitskiy} {and}
  \bibinfo{person}{Thomas Brox}.} \bibinfo{year}{2016}\natexlab{}.
\newblock \showarticletitle{Inverting Visual Representations with Convolutional
  Networks}. In \bibinfo{booktitle}{\emph{{IEEE} Conference on Computer Vision
  and Pattern Recognition}}.
\newblock


\bibitem[\protect\citeauthoryear{Du, Han, and Chen}{Du et~al\mbox{.}}{2004}]%
        {du2004privacy}
\bibfield{author}{\bibinfo{person}{Wenliang Du}, \bibinfo{person}{Yunghsiang~S
  Han}, {and} \bibinfo{person}{Shigang Chen}.} \bibinfo{year}{2004}\natexlab{}.
\newblock \showarticletitle{Privacy-preserving multivariate statistical
  analysis: Linear regression and classification}. In
  \bibinfo{booktitle}{\emph{Proceedings of the 2004 SIAM international
  conference on data mining}}.
\newblock


\bibitem[\protect\citeauthoryear{Fei-Fei~Li}{Fei-Fei~Li}{[n.d.]}]%
        {tinyimagenet}
\bibfield{author}{\bibinfo{person}{Justin~Johnson Fei-Fei~Li,
  Andrej~Karpathy}.} \bibinfo{year}{[n.d.]}\natexlab{}.
\newblock \bibinfo{title}{{Tiny ImageNet Visual Recognition Challenge}}.
\newblock \bibinfo{howpublished}{\url{https://tiny-imagenet.herokuapp.com/}}.
\newblock


\bibitem[\protect\citeauthoryear{Gentry}{Gentry}{2009}]%
        {gentry2009fully}
\bibfield{author}{\bibinfo{person}{Craig Gentry}.}
  \bibinfo{year}{2009}\natexlab{}.
\newblock \bibinfo{booktitle}{\emph{A fully homomorphic encryption scheme}}.
\newblock \bibinfo{publisher}{Stanford University}.
\newblock


\bibitem[\protect\citeauthoryear{Gilad{-}Bachrach, Dowlin, Laine, Lauter,
  Naehrig, and Wernsing}{Gilad{-}Bachrach et~al\mbox{.}}{2016}]%
        {gilad2016cryptonets}
\bibfield{author}{\bibinfo{person}{Ran Gilad{-}Bachrach},
  \bibinfo{person}{Nathan Dowlin}, \bibinfo{person}{Kim Laine},
  \bibinfo{person}{Kristin~E. Lauter}, \bibinfo{person}{Michael Naehrig}, {and}
  \bibinfo{person}{John Wernsing}.} \bibinfo{year}{2016}\natexlab{}.
\newblock \showarticletitle{CryptoNets: Applying Neural Networks to Encrypted
  Data with High Throughput and Accuracy}. In
  \bibinfo{booktitle}{\emph{Proceedings of the 33nd International Conference on
  Machine Learning,}}.
\newblock


\bibitem[\protect\citeauthoryear{Glorot and Bengio}{Glorot and Bengio}{2010}]%
        {glorot2010understanding}
\bibfield{author}{\bibinfo{person}{Xavier Glorot} {and} \bibinfo{person}{Yoshua
  Bengio}.} \bibinfo{year}{2010}\natexlab{}.
\newblock \showarticletitle{Understanding the difficulty of training deep
  feedforward neural networks}. In \bibinfo{booktitle}{\emph{Proceedings of the
  thirteenth international conference on artificial intelligence and
  statistics}}. \bibinfo{pages}{249--256}.
\newblock


\bibitem[\protect\citeauthoryear{Gu, Jamjoom, Su, Huang, Zhang, Ma, Pendarakis,
  and Molloy}{Gu et~al\mbox{.}}{2019}]%
        {gu2019reaching}
\bibfield{author}{\bibinfo{person}{Zhongshu Gu}, \bibinfo{person}{Hani
  Jamjoom}, \bibinfo{person}{Dong Su}, \bibinfo{person}{Heqing Huang},
  \bibinfo{person}{Jialong Zhang}, \bibinfo{person}{Tengfei Ma},
  \bibinfo{person}{Dimitrios Pendarakis}, {and} \bibinfo{person}{Ian Molloy}.}
  \bibinfo{year}{2019}\natexlab{}.
\newblock \showarticletitle{Reaching data confidentiality and model
  accountability on the caltrain}. In \bibinfo{booktitle}{\emph{2019 49th
  Annual IEEE/IFIP International Conference on Dependable Systems and Networks
  (DSN)}}. IEEE, \bibinfo{pages}{336--348}.
\newblock


\bibitem[\protect\citeauthoryear{Gueron}{Gueron}{2016}]%
        {gueron2016memory}
\bibfield{author}{\bibinfo{person}{Shay Gueron}.}
  \bibinfo{year}{2016}\natexlab{}.
\newblock \showarticletitle{A Memory Encryption Engine Suitable for General
  Purpose Processors}.
\newblock \bibinfo{journal}{\emph{{IACR} Cryptology ePrint Archive}}
  (\bibinfo{year}{2016}).
\newblock
\urldef\tempurl%
\url{http://eprint.iacr.org/2016/204}
\showURL{%
\tempurl}


\bibitem[\protect\citeauthoryear{Hanzlik, Zhang, Grosse, Salem, Augustin,
  Backes, and Fritz}{Hanzlik et~al\mbox{.}}{2018}]%
        {hanzlik2018mlcapsule}
\bibfield{author}{\bibinfo{person}{Lucjan Hanzlik}, \bibinfo{person}{Yang
  Zhang}, \bibinfo{person}{Kathrin Grosse}, \bibinfo{person}{Ahmed Salem},
  \bibinfo{person}{Max Augustin}, \bibinfo{person}{Michael Backes}, {and}
  \bibinfo{person}{Mario Fritz}.} \bibinfo{year}{2018}\natexlab{}.
\newblock \showarticletitle{Mlcapsule: Guarded offline deployment of machine
  learning as a service}.
\newblock \bibinfo{journal}{\emph{arXiv preprint arXiv:1808.00590}}
  (\bibinfo{year}{2018}).
\newblock


\bibitem[\protect\citeauthoryear{Huang, Liu, van~der Maaten, and
  Weinberger}{Huang et~al\mbox{.}}{2017}]%
        {huang2017densely}
\bibfield{author}{\bibinfo{person}{Gao Huang}, \bibinfo{person}{Zhuang Liu},
  \bibinfo{person}{Laurens van~der Maaten}, {and} \bibinfo{person}{Kilian~Q
  Weinberger}.} \bibinfo{year}{2017}\natexlab{}.
\newblock \showarticletitle{Densely connected convolutional networks}. In
  \bibinfo{booktitle}{\emph{Proceedings of the IEEE Conference on Computer
  Vision and Pattern Recognition}}.
\newblock


\bibitem[\protect\citeauthoryear{Hunt, Song, Shokri, Shmatikov, and
  Witchel}{Hunt et~al\mbox{.}}{2018}]%
        {hunt2018chiron}
\bibfield{author}{\bibinfo{person}{Tyler Hunt}, \bibinfo{person}{Congzheng
  Song}, \bibinfo{person}{Reza Shokri}, \bibinfo{person}{Vitaly Shmatikov},
  {and} \bibinfo{person}{Emmett Witchel}.} \bibinfo{year}{2018}\natexlab{}.
\newblock \showarticletitle{Chiron: Privacy-preserving Machine Learning as a
  Service}.
\newblock \bibinfo{journal}{\emph{arXiv preprint arXiv:1803.05961}}
  (\bibinfo{year}{2018}).
\newblock


\bibitem[\protect\citeauthoryear{Hynes, Cheng, and Song}{Hynes
  et~al\mbox{.}}{2018}]%
        {hynes2018efficient}
\bibfield{author}{\bibinfo{person}{Nick Hynes}, \bibinfo{person}{Raymond
  Cheng}, {and} \bibinfo{person}{Dawn Song}.} \bibinfo{year}{2018}\natexlab{}.
\newblock \showarticletitle{Efficient Deep Learning on Multi-Source Private
  Data}.
\newblock \bibinfo{journal}{\emph{arXiv preprint arXiv:1807.06689}}
  (\bibinfo{year}{2018}).
\newblock


\bibitem[\protect\citeauthoryear{Intel}{Intel}{2018}]%
        {sgxremoteattestation}
Intel \bibinfo{year}{2018}\natexlab{}.
\newblock \bibinfo{title}{{Intel Software Guard Extensions Remote Attestation
  End-to-End Example}}.
\newblock
  \bibinfo{howpublished}{\url{https://software.intel.com/en-us/articles/code-sample-intel-software-guard-extensions-remote-attestation-end-to-end-example}}.
\newblock


\bibitem[\protect\citeauthoryear{Jagannathan and Wright}{Jagannathan and
  Wright}{2005}]%
        {jagannathan2005privacy}
\bibfield{author}{\bibinfo{person}{Geetha Jagannathan} {and}
  \bibinfo{person}{Rebecca~N Wright}.} \bibinfo{year}{2005}\natexlab{}.
\newblock \showarticletitle{Privacy-preserving distributed k-means clustering
  over arbitrarily partitioned data}. In \bibinfo{booktitle}{\emph{Proceedings
  of the eleventh ACM SIGKDD international conference on Knowledge discovery in
  data mining}}.
\newblock


\bibitem[\protect\citeauthoryear{Kaplan, Powell, and Woller}{Kaplan
  et~al\mbox{.}}{2016}]%
        {kaplan2016amd}
\bibfield{author}{\bibinfo{person}{David Kaplan}, \bibinfo{person}{Jeremy
  Powell}, {and} \bibinfo{person}{Tom Woller}.}
  \bibinfo{year}{2016}\natexlab{}.
\newblock \bibinfo{booktitle}{\emph{{AMD Memory Encryption}}}.
\newblock \bibinfo{type}{{T}echnical {R}eport}. \bibinfo{institution}{White
  paper, AMD}.
\newblock


\bibitem[\protect\citeauthoryear{Krizhevsky, Sutskever, and Hinton}{Krizhevsky
  et~al\mbox{.}}{2012}]%
        {krizhevsky2012imagenet}
\bibfield{author}{\bibinfo{person}{Alex Krizhevsky}, \bibinfo{person}{Ilya
  Sutskever}, {and} \bibinfo{person}{Geoffrey~E. Hinton}.}
  \bibinfo{year}{2012}\natexlab{}.
\newblock \showarticletitle{ImageNet Classification with Deep Convolutional
  Neural Networks}. In \bibinfo{booktitle}{\emph{Advances in Neural Information
  Processing Systems}}.
\newblock


\bibitem[\protect\citeauthoryear{Li, Lai, Suda, Chandra, and Pan}{Li
  et~al\mbox{.}}{2017}]%
        {li2017privynet}
\bibfield{author}{\bibinfo{person}{Meng Li}, \bibinfo{person}{Liangzhen Lai},
  \bibinfo{person}{Naveen Suda}, \bibinfo{person}{Vikas Chandra}, {and}
  \bibinfo{person}{David~Z Pan}.} \bibinfo{year}{2017}\natexlab{}.
\newblock \showarticletitle{PrivyNet: A Flexible Framework for
  Privacy-Preserving Deep Neural Network Training with A Fine-Grained Privacy
  Control}.
\newblock \bibinfo{journal}{\emph{arXiv preprint arXiv:1709.06161}}
  (\bibinfo{year}{2017}).
\newblock


\bibitem[\protect\citeauthoryear{Lindell and Pinkas}{Lindell and
  Pinkas}{2000}]%
        {lindell2000privacy}
\bibfield{author}{\bibinfo{person}{Yehuda Lindell} {and} \bibinfo{person}{Benny
  Pinkas}.} \bibinfo{year}{2000}\natexlab{}.
\newblock \showarticletitle{Privacy preserving data mining}. In
  \bibinfo{booktitle}{\emph{Advances in Cryptology-CRYPTO 2000}}.
\newblock


\bibitem[\protect\citeauthoryear{Liu, Juuti, Lu, and Asokan}{Liu
  et~al\mbox{.}}{2017}]%
        {liu2017oblivious}
\bibfield{author}{\bibinfo{person}{Jian Liu}, \bibinfo{person}{Mika Juuti},
  \bibinfo{person}{Yao Lu}, {and} \bibinfo{person}{N Asokan}.}
  \bibinfo{year}{2017}\natexlab{}.
\newblock \showarticletitle{Oblivious Neural Network Predictions via MiniONN
  Transformations}. In \bibinfo{booktitle}{\emph{ACM Conference on Computer and
  Communications Security (CCS)}}.
\newblock


\bibitem[\protect\citeauthoryear{Mahendran and Vedaldi}{Mahendran and
  Vedaldi}{2015}]%
        {mahendran2015understanding}
\bibfield{author}{\bibinfo{person}{Aravindh Mahendran} {and}
  \bibinfo{person}{Andrea Vedaldi}.} \bibinfo{year}{2015}\natexlab{}.
\newblock \showarticletitle{Understanding deep image representations by
  inverting them}. In \bibinfo{booktitle}{\emph{{IEEE} Conference on Computer
  Vision and Pattern Recognition}}.
\newblock


\bibitem[\protect\citeauthoryear{McKeen, Alexandrovich, Anati, Caspi, Johnson,
  Leslie-Hurd, and Rozas}{McKeen et~al\mbox{.}}{2016}]%
        {mckeen2016intel}
\bibfield{author}{\bibinfo{person}{Frank McKeen}, \bibinfo{person}{Ilya
  Alexandrovich}, \bibinfo{person}{Ittai Anati}, \bibinfo{person}{Dror Caspi},
  \bibinfo{person}{Simon Johnson}, \bibinfo{person}{Rebekah Leslie-Hurd}, {and}
  \bibinfo{person}{Carlos Rozas}.} \bibinfo{year}{2016}\natexlab{}.
\newblock \showarticletitle{{Intel Software Guard Extensions (Intel SGX)
  Support for Dynamic Memory Management Inside an Enclave}}. In
  \bibinfo{booktitle}{\emph{Proceedings of the Hardware and Architectural
  Support for Security and Privacy 2016}} (Seoul, Republic of Korea)
  \emph{(\bibinfo{series}{HASP 2016})}. \bibinfo{publisher}{ACM},
  \bibinfo{address}{New York, NY, USA}, Article \bibinfo{articleno}{10},
  \bibinfo{numpages}{9}~pages.
\newblock
\showISBNx{978-1-4503-4769-3}
\urldef\tempurl%
\url{https://doi.org/10.1145/2948618.2954331}
\showDOI{\tempurl}


\bibitem[\protect\citeauthoryear{McKeen, Alexandrovich, Berenzon, Rozas, Shafi,
  Shanbhogue, and Savagaonkar}{McKeen et~al\mbox{.}}{2013}]%
        {mcKeen2013innovative}
\bibfield{author}{\bibinfo{person}{Frank McKeen}, \bibinfo{person}{Ilya
  Alexandrovich}, \bibinfo{person}{Alex Berenzon}, \bibinfo{person}{Carlos~V.
  Rozas}, \bibinfo{person}{Hisham Shafi}, \bibinfo{person}{Vedvyas Shanbhogue},
  {and} \bibinfo{person}{Uday~R. Savagaonkar}.}
  \bibinfo{year}{2013}\natexlab{}.
\newblock \showarticletitle{Innovative Instructions and Software Model for
  Isolated Execution}. In \bibinfo{booktitle}{\emph{Proceedings of the 2nd
  International Workshop on Hardware and Architectural Support for Security and
  Privacy}}.
\newblock


\bibitem[\protect\citeauthoryear{McMahan, Moore, Ramage, Hampson,
  et~al\mbox{.}}{McMahan et~al\mbox{.}}{2016}]%
        {mcmahan2016communication}
\bibfield{author}{\bibinfo{person}{H~Brendan McMahan}, \bibinfo{person}{Eider
  Moore}, \bibinfo{person}{Daniel Ramage}, \bibinfo{person}{Seth Hampson},
  {et~al\mbox{.}}} \bibinfo{year}{2016}\natexlab{}.
\newblock \showarticletitle{Communication-efficient learning of deep networks
  from decentralized data}.
\newblock \bibinfo{journal}{\emph{arXiv preprint arXiv:1602.05629}}
  (\bibinfo{year}{2016}).
\newblock


\bibitem[\protect\citeauthoryear{Mohassel and Zhang}{Mohassel and
  Zhang}{2017}]%
        {mohassel2017secureml}
\bibfield{author}{\bibinfo{person}{Payman Mohassel} {and}
  \bibinfo{person}{Yupeng Zhang}.} \bibinfo{year}{2017}\natexlab{}.
\newblock \showarticletitle{SecureML: A System for Scalable Privacy-Preserving
  Machine Learning.}. In \bibinfo{booktitle}{\emph{38th IEEE Symposium on
  Security and Privacy}}.
\newblock


\bibitem[\protect\citeauthoryear{Ohrimenko, Schuster, Fournet, Mehta, Nowozin,
  Vaswani, and Costa}{Ohrimenko et~al\mbox{.}}{2016}]%
        {ohrimenko2016oblivious}
\bibfield{author}{\bibinfo{person}{Olga Ohrimenko}, \bibinfo{person}{Felix
  Schuster}, \bibinfo{person}{C{\'e}dric Fournet}, \bibinfo{person}{Aastha
  Mehta}, \bibinfo{person}{Sebastian Nowozin}, \bibinfo{person}{Kapil Vaswani},
  {and} \bibinfo{person}{Manuel Costa}.} \bibinfo{year}{2016}\natexlab{}.
\newblock \showarticletitle{Oblivious Multi-Party Machine Learning on Trusted
  Processors.}. In \bibinfo{booktitle}{\emph{USENIX Security Symposium}}.
\newblock


\bibitem[\protect\citeauthoryear{Ossia, Shamsabadi, Taheri, Rabiee, Lane, and
  Haddadi}{Ossia et~al\mbox{.}}{2017}]%
        {ossia2017hybrid}
\bibfield{author}{\bibinfo{person}{Seyed~Ali Ossia},
  \bibinfo{person}{Ali~Shahin Shamsabadi}, \bibinfo{person}{Ali Taheri},
  \bibinfo{person}{Hamid~R Rabiee}, \bibinfo{person}{Nic Lane}, {and}
  \bibinfo{person}{Hamed Haddadi}.} \bibinfo{year}{2017}\natexlab{}.
\newblock \showarticletitle{A Hybrid Deep Learning Architecture for
  Privacy-Preserving Mobile Analytics}.
\newblock \bibinfo{journal}{\emph{arXiv preprint arXiv:1703.02952}}
  (\bibinfo{year}{2017}).
\newblock


\bibitem[\protect\citeauthoryear{Redmon}{Redmon}{2016}]%
        {darknet13}
\bibfield{author}{\bibinfo{person}{Joseph Redmon}.}
  \bibinfo{year}{2013--2016}\natexlab{}.
\newblock \bibinfo{title}{Darknet: Open Source Neural Networks in C}.
\newblock \bibinfo{howpublished}{\url{https://pjreddie.com/darknet/}}.
\newblock


\bibitem[\protect\citeauthoryear{Schwarz, Lipp, Moghimi, Van~Bulck, Stecklina,
  Prescher, and Gruss}{Schwarz et~al\mbox{.}}{2019}]%
        {schwarz2019zombieload}
\bibfield{author}{\bibinfo{person}{Michael Schwarz}, \bibinfo{person}{Moritz
  Lipp}, \bibinfo{person}{Daniel Moghimi}, \bibinfo{person}{Jo Van~Bulck},
  \bibinfo{person}{Julian Stecklina}, \bibinfo{person}{Thomas Prescher}, {and}
  \bibinfo{person}{Daniel Gruss}.} \bibinfo{year}{2019}\natexlab{}.
\newblock \showarticletitle{ZombieLoad: Cross-privilege-boundary data
  sampling}.
\newblock \bibinfo{journal}{\emph{arXiv preprint arXiv:1905.05726}}
  (\bibinfo{year}{2019}).
\newblock


\bibitem[\protect\citeauthoryear{Selvaraju, Cogswell, Das, Vedantam, Parikh,
  Batra, et~al\mbox{.}}{Selvaraju et~al\mbox{.}}{2017}]%
        {selvaraju2017grad}
\bibfield{author}{\bibinfo{person}{Ramprasaath~R Selvaraju},
  \bibinfo{person}{Michael Cogswell}, \bibinfo{person}{Abhishek Das},
  \bibinfo{person}{Ramakrishna Vedantam}, \bibinfo{person}{Devi Parikh},
  \bibinfo{person}{Dhruv Batra}, {et~al\mbox{.}}}
  \bibinfo{year}{2017}\natexlab{}.
\newblock \showarticletitle{Grad-CAM: Visual Explanations from Deep Networks
  via Gradient-Based Localization.}. In \bibinfo{booktitle}{\emph{ICCV}}.
  \bibinfo{pages}{618--626}.
\newblock


\bibitem[\protect\citeauthoryear{Shokri and Shmatikov}{Shokri and
  Shmatikov}{2015}]%
        {shokri2015privacy}
\bibfield{author}{\bibinfo{person}{Reza Shokri} {and} \bibinfo{person}{Vitaly
  Shmatikov}.} \bibinfo{year}{2015}\natexlab{}.
\newblock \showarticletitle{Privacy-preserving deep learning}. In
  \bibinfo{booktitle}{\emph{Proceedings of the 22nd ACM SIGSAC conference on
  computer and communications security}}. ACM.
\newblock


\bibitem[\protect\citeauthoryear{Shwartz-Ziv and Tishby}{Shwartz-Ziv and
  Tishby}{2017}]%
        {shwartz2017opening}
\bibfield{author}{\bibinfo{person}{Ravid Shwartz-Ziv} {and}
  \bibinfo{person}{Naftali Tishby}.} \bibinfo{year}{2017}\natexlab{}.
\newblock \showarticletitle{Opening the black box of deep neural networks via
  information}.
\newblock \bibinfo{journal}{\emph{arXiv preprint arXiv:1703.00810}}
  (\bibinfo{year}{2017}).
\newblock


\bibitem[\protect\citeauthoryear{Simonyan, Vedaldi, and Zisserman}{Simonyan
  et~al\mbox{.}}{2013}]%
        {simonyan2013deep}
\bibfield{author}{\bibinfo{person}{Karen Simonyan}, \bibinfo{person}{Andrea
  Vedaldi}, {and} \bibinfo{person}{Andrew Zisserman}.}
  \bibinfo{year}{2013}\natexlab{}.
\newblock \showarticletitle{Deep inside convolutional networks: Visualising
  image classification models and saliency maps}.
\newblock \bibinfo{journal}{\emph{arXiv preprint arXiv:1312.6034}}
  (\bibinfo{year}{2013}).
\newblock


\bibitem[\protect\citeauthoryear{Springenberg, Dosovitskiy, Brox, and
  Riedmiller}{Springenberg et~al\mbox{.}}{2014}]%
        {springenberg2014striving}
\bibfield{author}{\bibinfo{person}{Jost~Tobias Springenberg},
  \bibinfo{person}{Alexey Dosovitskiy}, \bibinfo{person}{Thomas Brox}, {and}
  \bibinfo{person}{Martin Riedmiller}.} \bibinfo{year}{2014}\natexlab{}.
\newblock \showarticletitle{Striving for simplicity: The all convolutional
  net}.
\newblock \bibinfo{journal}{\emph{arXiv preprint arXiv:1412.6806}}
  (\bibinfo{year}{2014}).
\newblock


\bibitem[\protect\citeauthoryear{Tramer and Boneh}{Tramer and Boneh}{2018}]%
        {tramer2018slalom}
\bibfield{author}{\bibinfo{person}{Florian Tramer} {and} \bibinfo{person}{Dan
  Boneh}.} \bibinfo{year}{2018}\natexlab{}.
\newblock \showarticletitle{Slalom: Fast, Verifiable and Private Execution of
  Neural Networks in Trusted Hardware}.
\newblock \bibinfo{journal}{\emph{arXiv preprint arXiv:1806.03287}}
  (\bibinfo{year}{2018}).
\newblock


\bibitem[\protect\citeauthoryear{Vaidya and Clifton}{Vaidya and
  Clifton}{2002}]%
        {vaidya2002privacy}
\bibfield{author}{\bibinfo{person}{Jaideep Vaidya} {and} \bibinfo{person}{Chris
  Clifton}.} \bibinfo{year}{2002}\natexlab{}.
\newblock \showarticletitle{Privacy preserving association rule mining in
  vertically partitioned data}. In \bibinfo{booktitle}{\emph{Proceedings of the
  eighth ACM SIGKDD international conference on Knowledge discovery and data
  mining}}.
\newblock


\bibitem[\protect\citeauthoryear{Vaidya, Kantarc{\i}o{\u{g}}lu, and
  Clifton}{Vaidya et~al\mbox{.}}{2008}]%
        {vaidya2008privacy}
\bibfield{author}{\bibinfo{person}{Jaideep Vaidya}, \bibinfo{person}{Murat
  Kantarc{\i}o{\u{g}}lu}, {and} \bibinfo{person}{Chris Clifton}.}
  \bibinfo{year}{2008}\natexlab{}.
\newblock \showarticletitle{Privacy-preserving naive bayes classification}.
\newblock \bibinfo{journal}{\emph{The VLDB Journal—The International Journal
  on Very Large Data Bases}} (\bibinfo{year}{2008}).
\newblock


\bibitem[\protect\citeauthoryear{Van~Bulck, Minkin, Weisse, Genkin, Kasikci,
  Piessens, Silberstein, Wenisch, Yarom, and Strackx}{Van~Bulck
  et~al\mbox{.}}{2018}]%
        {vanbulck2018foreshadow}
\bibfield{author}{\bibinfo{person}{Jo Van~Bulck}, \bibinfo{person}{Marina
  Minkin}, \bibinfo{person}{Ofir Weisse}, \bibinfo{person}{Daniel Genkin},
  \bibinfo{person}{Baris Kasikci}, \bibinfo{person}{Frank Piessens},
  \bibinfo{person}{Mark Silberstein}, \bibinfo{person}{Thomas~F. Wenisch},
  \bibinfo{person}{Yuval Yarom}, {and} \bibinfo{person}{Raoul Strackx}.}
  \bibinfo{year}{2018}\natexlab{}.
\newblock \showarticletitle{Foreshadow: Extracting the Keys to the {Intel SGX}
  Kingdom with Transient Out-of-Order Execution}. In
  \bibinfo{booktitle}{\emph{Proceedings of the 27th {USENIX} Security
  Symposium}}. \bibinfo{publisher}{{USENIX} Association}.
\newblock


\bibitem[\protect\citeauthoryear{van Schaik, Milburn, {\"O}sterlund, Frigo,
  Maisuradze, Razavi, Bos, and Giuffrida}{van Schaik et~al\mbox{.}}{2019}]%
        {van2019ridl}
\bibfield{author}{\bibinfo{person}{Stephan van Schaik}, \bibinfo{person}{Alyssa
  Milburn}, \bibinfo{person}{Sebastian {\"O}sterlund}, \bibinfo{person}{Pietro
  Frigo}, \bibinfo{person}{Giorgi Maisuradze}, \bibinfo{person}{Kaveh Razavi},
  \bibinfo{person}{Herbert Bos}, {and} \bibinfo{person}{Cristiano Giuffrida}.}
  \bibinfo{year}{2019}\natexlab{}.
\newblock \showarticletitle{RIDL: Rogue In-Flight Data Load}.
\newblock \bibinfo{journal}{\emph{S\&P (May 2019)}} (\bibinfo{year}{2019}).
\newblock


\bibitem[\protect\citeauthoryear{Yosinski, Clune, Nguyen, Fuchs, and
  Lipson}{Yosinski et~al\mbox{.}}{2015}]%
        {yosinski2015understanding}
\bibfield{author}{\bibinfo{person}{Jason Yosinski}, \bibinfo{person}{Jeff
  Clune}, \bibinfo{person}{Anh~Mai Nguyen}, \bibinfo{person}{Thomas~J. Fuchs},
  {and} \bibinfo{person}{Hod Lipson}.} \bibinfo{year}{2015}\natexlab{}.
\newblock \showarticletitle{Understanding Neural Networks Through Deep
  Visualization}.
\newblock \bibinfo{journal}{\emph{CoRR}}  \bibinfo{volume}{abs/1506.06579}
  (\bibinfo{year}{2015}).
\newblock
\urldef\tempurl%
\url{http://arxiv.org/abs/1506.06579}
\showURL{%
\tempurl}


\bibitem[\protect\citeauthoryear{Zeiler and Fergus}{Zeiler and Fergus}{2014}]%
        {zeiler2014visualizing}
\bibfield{author}{\bibinfo{person}{Matthew~D. Zeiler} {and}
  \bibinfo{person}{Rob Fergus}.} \bibinfo{year}{2014}\natexlab{}.
\newblock \showarticletitle{{Visualizing and Understanding Convolutional
  Networks}}. In \bibinfo{booktitle}{\emph{Computer Vision - {ECCV} 2014 - 13th
  European Conference}}.
\newblock


\bibitem[\protect\citeauthoryear{Zhang}{Zhang}{2018}]%
        {mbedtls-SGX}
\bibfield{author}{\bibinfo{person}{Fan Zhang}.}
  \bibinfo{year}{2018}\natexlab{}.
\newblock \bibinfo{title}{{TLS for SGX: a port of mbedtls}}.
\newblock
  \bibinfo{howpublished}{\url{https://github.com/bl4ck5un/mbedtls-SGX}}.
\newblock


\bibitem[\protect\citeauthoryear{Zhou, Khosla, Lapedriza, Oliva, and
  Torralba}{Zhou et~al\mbox{.}}{2016}]%
        {zhou2016learning}
\bibfield{author}{\bibinfo{person}{Bolei Zhou}, \bibinfo{person}{Aditya
  Khosla}, \bibinfo{person}{Agata Lapedriza}, \bibinfo{person}{Aude Oliva},
  {and} \bibinfo{person}{Antonio Torralba}.} \bibinfo{year}{2016}\natexlab{}.
\newblock \showarticletitle{Learning deep features for discriminative
  localization}. In \bibinfo{booktitle}{\emph{Proceedings of the IEEE
  Conference on Computer Vision and Pattern Recognition}}.
  \bibinfo{pages}{2921--2929}.
\newblock


\end{thebibliography}
